\documentclass[12pt]{article}
\usepackage{latexsym,epsfig,amssymb,amsmath,subfigure,color,pictex,fullpage,amsfonts,ulem}

\usepackage{amssymb}
\usepackage{authblk}

\usepackage{lineno}

\usepackage{comment}
\usepackage{algorithm}
\usepackage[noend]{algpseudocode}
\makeatletter
\def\BState{\State\hskip-\ALG@thistlm}
\makeatother

\usepackage{latexsym}
\usepackage{epsfig}
\usepackage{amsmath}
\usepackage{subfigure}
\usepackage{pictex}
\usepackage{fullpage}
\usepackage{amsfonts}
\usepackage{float}
\usepackage[colorlinks=true]{hyperref}
\usepackage{placeins}

\restylefloat{table}

\usepackage{xcolor}

\newcommand{\ds}{\displaystyle}

\newcommand{\dd}{\partial}

\newcommand{\ueq}{U_{\rm eq}}

\newcommand{\B}{\mathcal{B}}

\newcommand{\tp}{t_0^+}
\newcommand{\De}{D_{\rm{eff}}}
\newcommand{\be}{\begin{eqnarray}}
\newcommand{\ee}{\end{eqnarray}}
\newcommand{\bes}{\begin{eqnarray*}}
	\newcommand{\ees}{\end{eqnarray*}}

\newcommand{\Fms}{{N}_{-}}

\newcommand{\phie}{\Phi}

\newcommand{\je}{j}

\newcommand{\js}{j_s}

\newcommand{\jn}{j_n}

\newcommand{\sigs}{\sigma_s}

\newcommand{\cs}{c_{s}}
\newcommand{\Ds}{D_{s}}
\newcommand{\ueqs}{U_{eq}(\cs|_{r=R})}
\newcommand{\phis}{\Phi_s}

\newcommand{\csm}{c_s^{\rm max}}

\def\bP{{\mathbf{P}}}
\def\J{{\mathcal{J}}}
\def\tV{{\widetilde{V}}}

\DeclareMathOperator{\arcsinh}{arcsinh}
\DeclareMathOperator{\rand}{rand}

\usepackage[margin=1in]{geometry}

\title{ {On Uncertainty Quantification in the Parametrization of
    Newman-type Models of Lithium-ion Batteries} }
\author[1,2]{Jose Morales Escalante}
\author[3,4]{Smita Sahu}
\author[3,4]{Jamie M. Foster}
\author[2]{Bartosz Protas\thanks{Corresponding Author, Email: {\tt bprotas@mcmaster.ca}}}
\affil[1]{Department of Mathematics, University of Texas at San Antonio, One UTSA Circle, San Antonio, Texas 78249}
\affil[2]{Department of Mathematics and Statistics, McMaster University
	Hamilton, Ontario, CANADA L8S 4K1}
\affil[3]{The Faraday Institution, Quad One, Becquerel Avenue, Harwell Campus, Didcot, OX11 0RA, UK}
\affil[4]{School of Mathematics and Physics, University of Portsmouth, Lion Terrace, PO1 3HF, UK}

\begin{document}
\maketitle

\begin{abstract}
  We consider the problem of parameterizing Newman-type models of
  Li-ion batteries focusing on quantifying the inherent uncertainty of
  this process and its dependence on the discharge rate. In order to
  rule out genuine experimental error and instead isolate the
  intrinsic uncertainty of model fitting, we concentrate on an
  idealized setting where ``synthetic'' measurements in the form of
  voltage curves are manufactured using the full, and most accurate,
  Newman model with parameter values considered ``true'', whereas
  parameterization is performed using simplified versions of the
  model, namely, the single-particle model and its recently proposed
  corrected version.  By framing the problem in this way, we are able
  to eliminate aspects which affect uncertainty, but are hard to
  quantity such as, e.g., experimental errors. The parameterization is
  performed by formulating an inverse problem which is solved using a
  state-of-the-art Bayesian approach in which the parameters to be
  inferred are represented {in terms of} suitable probability
  distributions; this allows us to assess the uncertainty of their
  reconstruction. {The key finding is that while at slow discharge
    rates the voltage curves can be reconstructed quite accurately,
    this can be achieved with some parameter varying by 300\% or more,
    thus providing evidence for very high uncertainty of the parameter
    inference process. As the discharge rate increases, the
    reconstruction uncertainty is reduced {but at the same time
      the fits to the voltage curves becomes less accurate}. These
    observations highlight the ill-posedness of the inverse problem of
    parameter reconstruction in models of Li-ion battery operation.}
  In practice, using simplified model appears to be a viable and
  useful strategy provided that the assumptions facilitating the model
  simplification are truly valid for the battery operating regimes in
  which the data was collected.
\end{abstract}
	
\section{Introduction}	

Lithium-ion batteries (LiBs) are already produced in their billions
each year and the industry is set to grow significantly over the
coming decades as electric vehicles rise in prevalence
\cite{Blo16,Che19}.  Even though the technology is already widespread,
improvements in cell lifetime, recyclability, and increased charging
rates are required. This development can be accelerated by
supplementing practical work with accurate models allowing in-silico
testing of novel designs without the need for costly and
time-consuming physical prototyping.

The Doyle-Fuller-Newman {(DFN)} model is the most-commonly used
modelling framework for describing the operation of LiBs on the scale
of anode-cathode pairs, i.e., on the cell level
\cite{doyle96,Doy93,Ful94,srinivasan}. Its utility has been
demonstrated many times, e.g., in \cite{Har19,Jin18,Krach,Zul21}, and
it has been systematically derived and analyzed in numerous other
works, see, e.g., \cite{ciucci,Ric12,Fos15,franco13,Jok16}. The model
is physics-based, and {has enough fidelity} to predict the
changes in performance that result from alterations in meaningful
device parameters, yet it is also coarse enough that it can be solved
on feasible timescales using relatively modest computational
{resources}. However, a persistent difficulty lies in developing
robust techniques to accurately estimate the large number of scalar
{material properties and state-dependent constitutive relations} that
are needed to parameterize it. Depending on which variant of the DFN
model is being used, typical numbers of parameters are in the range of
15--30.  Experimental methods can used to measure many of the required
parameters {(within the limits of experimental error)} and recent
years have seen examples of studies that have been able to obtain
complete parameterization for certain cells \cite{Ecker2015a,
    Ecker2015b, Schmalstieg2018a, Schmalstieg2018b}. Despite this,
complete experimental characterization remains difficult, time
consuming and requires {specialized} equipment.


Inverse modelling is an approach that can be used in parallel with
experimental characterization to help obtain parameters that cannot be
readily measured. The basic idea is to {combine measurement data with
  mathematical models of the processes to infer unknown parameters,
  typically by minimizing the discrepancy between the measured
  quantities and the corresponding quantities predicted by the model
  using methods of numerical optimization. One type of electrochemical
  data which has been often used for this purpose because it is
  relatively easy to obtain are discharge voltage curves.}  The
efficacy of the approach in battery modelling has already been
demonstrated a number of times, see e.g.,
{\cite{kgnhllf12,skhgp15a,rfskhgp17},} but the inverse modelling
approach becomes increasingly harder to apply as the number of model
parameters increases.

The main difficulty is that inverse problems tend to be {\it
  ill-posed}, in the sense that they usually do not admit exact
solutions in the form of parameters such that the corresponding model
predictions would match the measurements {\it exactly}. On the other
hand, inverse problems typically admit many, often infinitely-many,
approximate solutions where the model predictions corresponding to the
inferred parameters match the measurements only approximately.  This
ill-posedness has roots in weak dependence of the model predictions on
some of its parameters and is compounded by experimental noise and
numerical errors unavoidably present in approximations of the model
equations and in the solution of the optimization problem. Faced with
a multitude of possible approximate solutions, it is important to
assign relative uncertainties to each of them. However, this is
difficult when the inverse problem is formulated in the classical way
by defining the error functional and then minimizing it with respect
to unknown parameters using methods of numerical optimization, a
procedure which normally yields one set of ``optimal'' parameters. On
the other hand, uncertainty of reconstructed parameters can be
conveniently characterized in the framework of Bayesian inference
which recently began to attract a lot of attention
\cite{10.5555/2568154,Tenorio,KaipioSomersalo}. The main idea is to
frame the inverse problem in probabilistic terms, such that unknown
parameters are inferred in terms of their probability distributions.
Recent applications of this approach to problems in electrochemistry
are described in \cite{SethurajanBayesian,sethurajan19,amah20}.

The use of more, or higher-fidelity, experimental data can help
mitigate {the problem described above}, but another complimentary
approach is to write down simplified models, which contain fewer
parameters and to fit those instead. This latter strategy comes with
{the} caveat that one must take care that the simplified model is
still complete enough that it accurately captures the important
physical/chemical processes within the LiB. {Thus, there are subtle
  trade-offs between the fidelity of models, their parametrizability
  and their {subsequent utility}.}

The main goal of this paper is to demonstrate, for what we believe to
be the first time in the electrochemical literature, that even in
simple settings inverse modelling can in fact lead to ambiguous
results. This will be done by analyzing the uncertainty of
parameterization of relatively simple models introduced below using
methods of Bayesian inference and ``measurements'' obtained with the
DFN model. By framing the problem in this way we eliminate
experimental measurement errors and ensure all numerical errors are
strictly controlled, such that we can focus on the effect of
ill-posedness.

To fix attention, we focus on modelling a cathode made from nickel
manganese cobalt oxide Li(Ni$_{0.4}$Co$_{0.6}$)O$_2$, a material also
known as NMC or LNC whose properties were studied in \cite{Ecker2015a,
  Ecker2015b}. {In fact, the set-up of our problem can be
  regarded as a highly simplified ``in silico'' version of these
  experiments.}  In the next section we describe the hierarchy of
models that will be used in this study, first outlining the DFN model
and then describing the Single Particle Model (SP) and its corrected
variant (cSP). Next, in Section \ref{sec:inverse} we describe the
Bayesian approach to inverse modelling with uncertainty
quantification. Computational results are presented in Section
\ref{sec:results} whereas final conclusions and outlook are deferred
to Section \ref{sec:final}. Some additional technical material is
collected in Appendix \ref{sec:numerics}.

\section{Electrochemical models}
\label{sec:models} 

In this section we {introduce a hierarchy of models describing
  transport of electrochemically active species in a cell. We will
  focus on the galvanostatic set-up where the applied external current
  $I(t)$ is considered as the input and the drop of the voltage across
  the cell $V(t)$ is the main output.}  First, we {present} the
classical DFN model also known as the pseudo 2D (P2D) model. Developed
by Newman and his co-workers in the mid-’90s and early 2000’s
\cite{doyle96, Doy93,Ful94}, the DFN model is based on transport
equations for lithium ions in the electrolyte and {for lithium atoms}
in the active particles of the electrodes (the cathode and the anode)
{coupled through the Butler-Volmer relations describing the
  intercalation and de-intercalation of lithium at the interface
  between active particles and the electrolyte. The DFN model has the
  form of a parabolic-elliptic system of partial differential
  equations (PDEs) which after discretization in space leads to a
  system of differential-algebraic equations (DAEs). This model is
  thus computationally complex which limits its use in some
  applications.}

The complexity of the DFN model is the motivation behind the
{development of} reduced-order models. One {such} model is the
{single-particle} (SP) model in which it is assumed that particles
throughout the electrode thickness behave in the same way such that
the transport in only a single particle needs to be solved for.
Systematic derivations of the SP model from the DFN model have been
given in \cite{Marquis19,Richardson20} and other justifications have
been given in \cite{Mou16,Guo,Kemper13}.

{The corrected single-particle (cSP) model is derived using a formal
  asymptotic method applied to the DFN model \cite{Richardson20}. It
  is based} on the disparity between the size of thermal voltage and
of the characteristic change in overpotential that occurs during
(de)lithiation. It has the advantage that it more accurately
reproduces the voltage predicted by the DFN model than the SP model
but is slightly more expensive to solve (though still markedly simpler
than the full DFN model. {The two reduced-order models are related
  since} the asymptotic limit of large changes in the open-circuit
voltage (OCV) relative to the thermal voltage {recovers a} variant of
the SP model at leading order because the reaction overpotentials are
small.

\begin{figure}
  \centering
  \hspace*{1cm}
  \includegraphics[width=12cm]{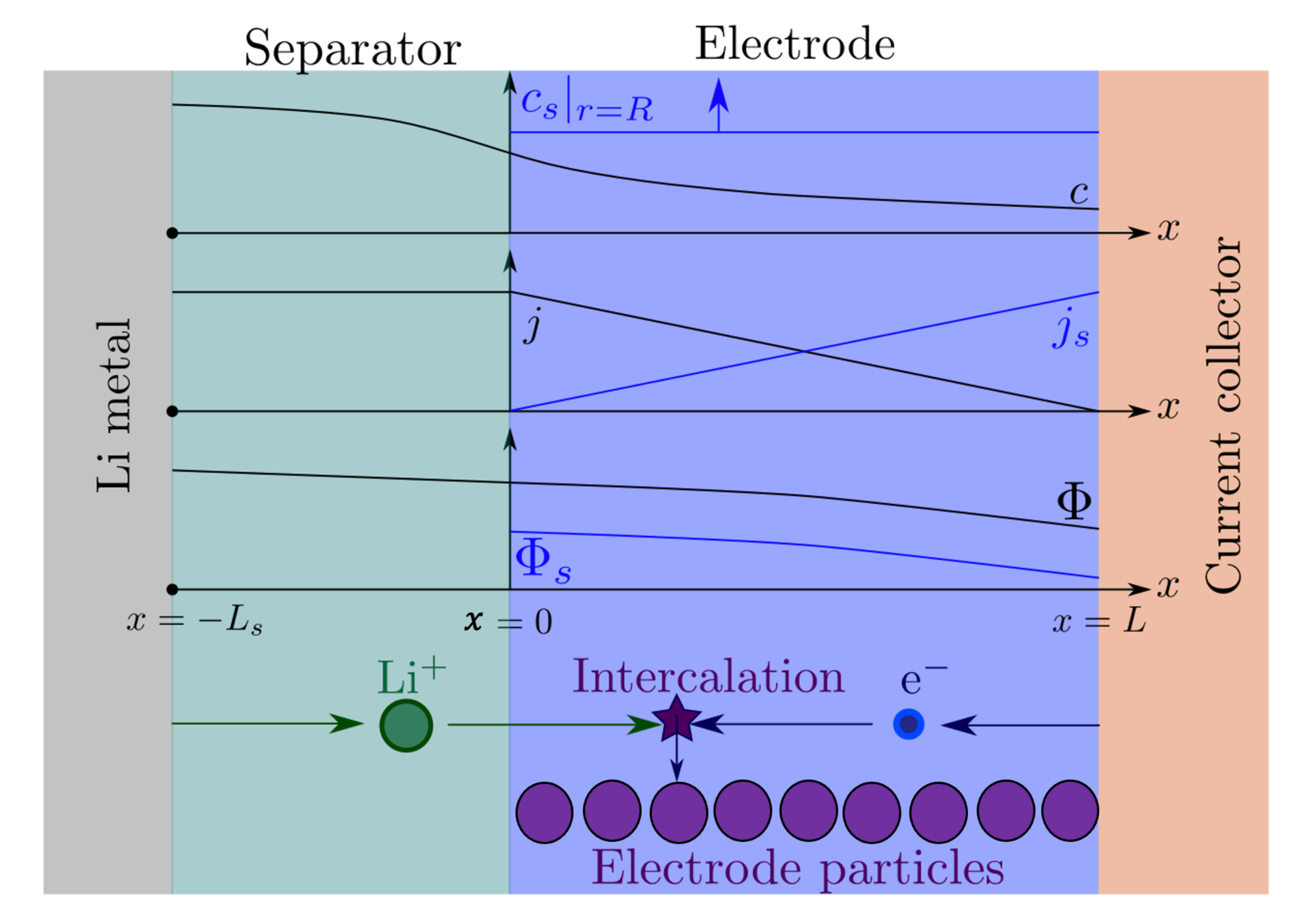}\hspace*{1cm}
  \caption{A schematic diagram of the half-cell geometry. The anode is
    made from lithium metal, the separator is located between $x=-L_s$
    and $x=0$, and the cathode current collector is at $x = L$.}
    \label{fig:halfcell}
\end{figure}

For simplicity of presentation, we will focus here on the half-cell
geometry illustrated schematically in Figure \ref{fig:halfcell} where
the coordinate $x$ measures distance across the cell and $r$ is the
radial coordinate in the active particles.

\subsection{Doyle-Fuller-Newman (DFN) model}
\label{sec:DFN}
We now briefly state the DFN model.

\paragraph{Macroscopic equations.}
The equations governing ionic transport through the electrolyte and
those governing electronic transport through the porous binder matrix
are
\begin{align}
\epsilon_l\frac{\partial c}{\partial t}+\frac{\partial \Fms}{\partial x}=0,\quad 
\Fms=-\B\De(c)\frac{\partial c}{\partial x}-(1-\tp)\frac{\je}{F} \quad \mbox{in} \quad -L_s < x < L,\label{hc1}\\
\frac{\partial \je}{\partial x}= b \jn, \qquad 
\je=-\B\kappa(c)\left(\frac{\partial \phie}{\partial x}-\frac{2R_gT}{F}\frac{1-\tp}{c}\frac{\partial c}{\partial x}\right)\quad \mbox{in} \quad L_s < x< L,\label{hc2}\\
\frac{\partial \js}{\partial x}=-b\jn,\quad 
\js=-{\sigs}\frac{\partial \phis}{\partial x} \quad \mbox{in} \quad 0< x<L, \label{hc3}\\
\jn=\left\{\begin{array}{ccc}
0 & \mbox{in} & -L_s < x < 0, \\
\ds 2 F k c^{1/2} \left( \cs \rvert _{r=R} \right)^{1/2} \left( \csm -\cs \rvert_{r=R}\right)^{1/2} \sinh\left(\frac{F\eta}{2R_gT}\right) &  \mbox{in} & 0 \leq x < L, \end{array} \right.\label{hc4}\\
\eta=\phis-\phie-\ueqs.\label{hc5} 
\end{align}
Here $x$ and $t$ denote position through the electrode and time
respectively, $L$ and $L_s$ are the widths of the separator and
electrode respectively, $\epsilon_l$ is the volume fraction of active
material, $c$ is the molar concentration of lithium ions in the
electrolyte, $\Fms$ is the effective flux of negative counterion
across the electrolyte, $\B$ is the permeability factor, $\De$ is the
ionic diffusivity, $\tp$ is the transference number, $\je$ and $\js$
are the ionic and electronic current densities, $F$ is Faraday's
constant, $R_g$ is the universal gas constant, $T$ is the absolute
temperature, $b$ is the Brunauer-Emmett-Teller (BET) surface area,
$\kappa$ is the ionic conductivity, $\jn$ is the current density
carried by the Butler-Volmer reaction, $\Phi$, $\phis$ are the
electric potential in the electrolyte and electrode respectively,
$\sigs$ is the conductivity of the electrode and $\ueq ,\csm$ and $R$
are the equilibrium potential, the maximum concentration that can be
stored in the electrode material, and {the radius of the electrode
  particles located at the macroscopic coordinate $x$.}

\paragraph{Macroscopic boundary conditions.} We impose {boundary
  conditions requiring that:} (i) the Li-metal counter-electrode
supplies an ionic current density of $\frac{I(t)}{A}$ to the
electrolyte on $x=-L_s$; (ii) current of density $\frac{I(t)}{A}$ is
extracted from the solid phase of the electrode at the current
collector $x=L$; and (iii) {at} $x=0$ no current flows from the solid
phase into the separator. In addition, (iv) the potential $\Phi$ in
the electrolyte is specified on $x=-L_s$ (edge of the separator) and
(v) and (vi) there is no counterion flux across the interfaces between
the electrolyte and the separator and between the electrolyte and the
current collector.  In summary, this gives, cf.~Figure
\ref{fig:halfcell},
\begin{align}
  &\je|_{x=-L_s}= \frac{I}{A},\quad \Fms|_{x=-L_s}= 0,\quad \Phi|_{x=-L_s}= 0,\quad \Fms|_{x=L}= 0.\label{hc6}\\
  &\js|_{x=0}= 0,\quad \js|_{x=L}= \frac{I}{A}.\label{hc7}
 \end{align}
 
\paragraph{Microscopic equations and boundary conditions.}
Conservation of lithium in a single spherical active material particle
of radius $R$ is described by the diffusion equation {with
  suitable boundary conditions} 
\be
\frac{\dd \cs}{\dd t}=\frac{1}{r^{2}}\frac{\dd }{\dd r}\left( r^{2} \Ds (\cs)\frac{\dd \cs}{\dd r}\right) \quad \mbox{in} \quad 0< r<R,\label{hc8}\\
\cs \,\, \mbox{bounded} \,\, \mbox{on} \,\, r=0, \qquad -\Ds(\cs)
\frac{\partial \cs}{\partial r}\bigg\rvert_{r= R}=\frac{\jn}{F}\
\ \mbox{in} \ \ 0< x<L.\label{hc9} \ee

\paragraph{Initial conditions.} Furthermore, partial differential
equations from {the system} \eqref{hc1}--\eqref{hc9} require
initial conditions on concentration in electrolyte $c$ and in
electrode $c_s$. Initially, we assume that electrode is in a state of
equilibrium, i.e., $I(0)=0$, and $j_n(x,0)=0$ {then} gives the
uniform electrolyte salt concentration
\be\label{hc11}
c|_{t=0}= c_{init}\quad  \mbox{in} \quad -L_s < x < L,
\ee
and a uniform lithium concentration in all the electrode particles
\be\label{hc12}
c_s|_{t=0}=c_{s,init}\quad\mbox{in} \quad 0<r<R.
\ee
These conditions together imply that $\Phi(x,0)=0$ and $\Phi_s(x,0)=\ueq(c_{s,init})$.

\paragraph{Geometry.} We assume that all electrode particles are
spherical {such that} the BET surface area $b$ and electrode
particle volume fraction $\epsilon_s$ are related by the expression
\be\label{eplislons}
\epsilon_s=\frac{bR}{3}.
\ee 
Furthermore, the electrolyte volume fraction $\epsilon_l$ is related to $\epsilon_s$ by
\be\label{epsilonl}
\epsilon_l = 1- \epsilon_{inert} - \epsilon_s,
\ee
where $\epsilon_{inert}$ is the volume fraction of electrochemically inert material.

\paragraph{The half-cell potential.}  Once the model equations have
been solved, the half-cell potential $V$ can be evaluated using
\be\label{hc10}
V(t)= \Phi_s|_{x=L}.
\ee

\paragraph{C-rate.} 
It is common to quantify the current being drawn from/supplied to a
cell using a measure called the C-rate; this is a dimensionless
quantity defined to be the present current supply/draw divided by the
current supply/draw that is needed to ``fully (dis)charge'' the cell
in one hour. In the interests of consistency, we shall define our
C-rate in the same fashion as was done by Ecker et.~al.~in
\cite{Ecker2015a,Ecker2015b} who experimentally determined that a
current of 0.15625A ``fully utilized'' their LNC cathode in 1 hour.
Our C-rate is then defined as
\begin{equation}
\text{C-rate} = \frac{I}{0.15625}
\end{equation}
{and can be changed by varying the current $I$.} 

We note that there is a small discrepancy between the cell capacity
implied by Ecker et.~al.'s experimental measurement and the theoretical
capacity (which is marginally higher). An upshot of this is that
later, see e.g., Figure \ref{fig:V}, we are able to maintain cell
operation for slightly longer than expected.

\subsection{Single-particle (SP) model}
\label{sec:SPM}
{As the most common
  approximation to the DFN model \eqref{hc1}--\eqref{hc12}, the SP
  model is obtained by assuming that the cell voltage only depends} upon
the potential of the insertion material. To determine this
{quantity} we solve only for transport in a single representative
insertion particle. All other potential drops are neglected, including
that across the electrolyte, across the solid, and across the
interfaces (double layers) between the electrolyte and insertion
material caused the interfacial resistance. There are different
instances when this level of simplification is justified, for example,
during low C-rate operation. {The SP model is then given by}
\begin{align}
&\frac{\dd \cs}{\dd t}=\frac{1}{r^{2}}\frac{\dd }{\dd r}\left( r^{2} \Ds (\cs)\frac{\dd \cs}{\dd r}\right) \quad \mbox{in} \quad 0< r<R,\label{spm1}\\
&\cs\,\, \mbox{bounded} \,\, \mbox{on} \,\, r=0, \qquad \cs|_{r=R}= \mathcal{C}(t),\qquad \cs|_{t=0}=c_{s,init},\label{spm2}\\
&\int_0^L j_n(x,t)dx = -\frac{I(t)}{Ab},\,\, \mbox{where} \,\,  j_n =-F\Ds(\cs) \frac{\partial \cs}{\partial r}\bigg\rvert_{r= R}\label{spm3},
\end{align}
where lithium concentration on the surface of the electrode particle
is $\mathcal{C}(t)$ and the {relation \eqref{spm3} implies that the
  total reaction {output} is equal to the net current $I(t)$ thus
  ensuring the conservation of charge.} The voltage $V(t)$ of the
half-cell is then estimated using
\begin{equation}
\label{spm4}
V(t)=\ueq\left(\mathcal{C}(t)\right).
\end{equation}
A systematic derivation of the SP model from the DFN model based on
the asymptotic limit of large electrode and electrolyte conductivities
and large electrolyte diffusivity is presented in \cite{Marquis19},
whereas an extension of this model is introduced in \cite{Kemper13}.
Parameter and state estimation problems for the SP model were recently
considered in \cite{Mou16,bizeray18}.  {{Similar asymptotic techniques
    for LiB in the context of a porous electrode model have been used
    in \cite{moyles18}.  }}

\subsection{Corrected single-particle (cSP) model}
\label{sec:cSPM}

The corrected single-particle (cSP) model introduced in
\cite{Richardson20} is accurate for materials that have an
overpotential which varies appreciably with concentration (in
practice, {these are } most common electrode materials, except for
lithium-iron-phosphate) and for sufficiently low C-rates.  However, it
is worth emphasizing that, as demonstrated in \cite{Richardson20}, the
cSP {model} remains accurate for much larger C-rates than the SP
model.

The cSP model consists of the standard SP model
\eqref{spm1}--\eqref{spm4} supplemented by the governing equations for
the electrolyte concentration and potential given in
\eqref{hc1}--\eqref{hc2} which are subject to the boundary and initial
condition
\begin{align}
&\je|_{x=-L_s}= \frac{I(t)}{A},\quad \Fms|_{x=-L_s}= 0\quad \Phi|_{x=-L_s}= 0,\quad \Fms|_{x=L}= 0, \label{ele_bc}\\ 
&c|_{t=0}=c_{init}. \label{ele_ic}
\end{align}
The model equations for the SP model are decoupled from those for the
electrolyte and can hence be solved first. Subsequently, the evolution
of {the concentration and potential in} the electrolyte is found
and, at this stage, the Butler-Volmer current density is known from
(\ref{hc9}b). Finally, we calculate the corrected half-cell voltage by
evaluating
\be
\label{c_voltage} V(t) = \ueq(\mathcal{C}(t)) +\frac{\int_0^L \left[ \eta(x,t) +\phie(x,t) - \int_0^L \frac{\js(x',t)dx'}{\sigma_s}\right]dx}{ L},
\ee
where the current density is given by
\begin{align} 
&\js(x,t)=\frac{I}{A}-j(x,t)
\end{align}
and overpotential is
\begin{align} 
&\eta(x,t) = 2\frac{R_gT}{F}\arcsinh\left( \frac{j_n(x,t)}{2Fk\left[(\csm-\mathcal{C}(t))\mathcal{C}(t)c\right]^{1/2}}\right). \label{phi_solid}
\end{align}

\subsection{Test Problems}
\label{sec:problem}

\begin{figure}
  \centering
  \mbox{
    \subfigure[]{\includegraphics[width=7cm]{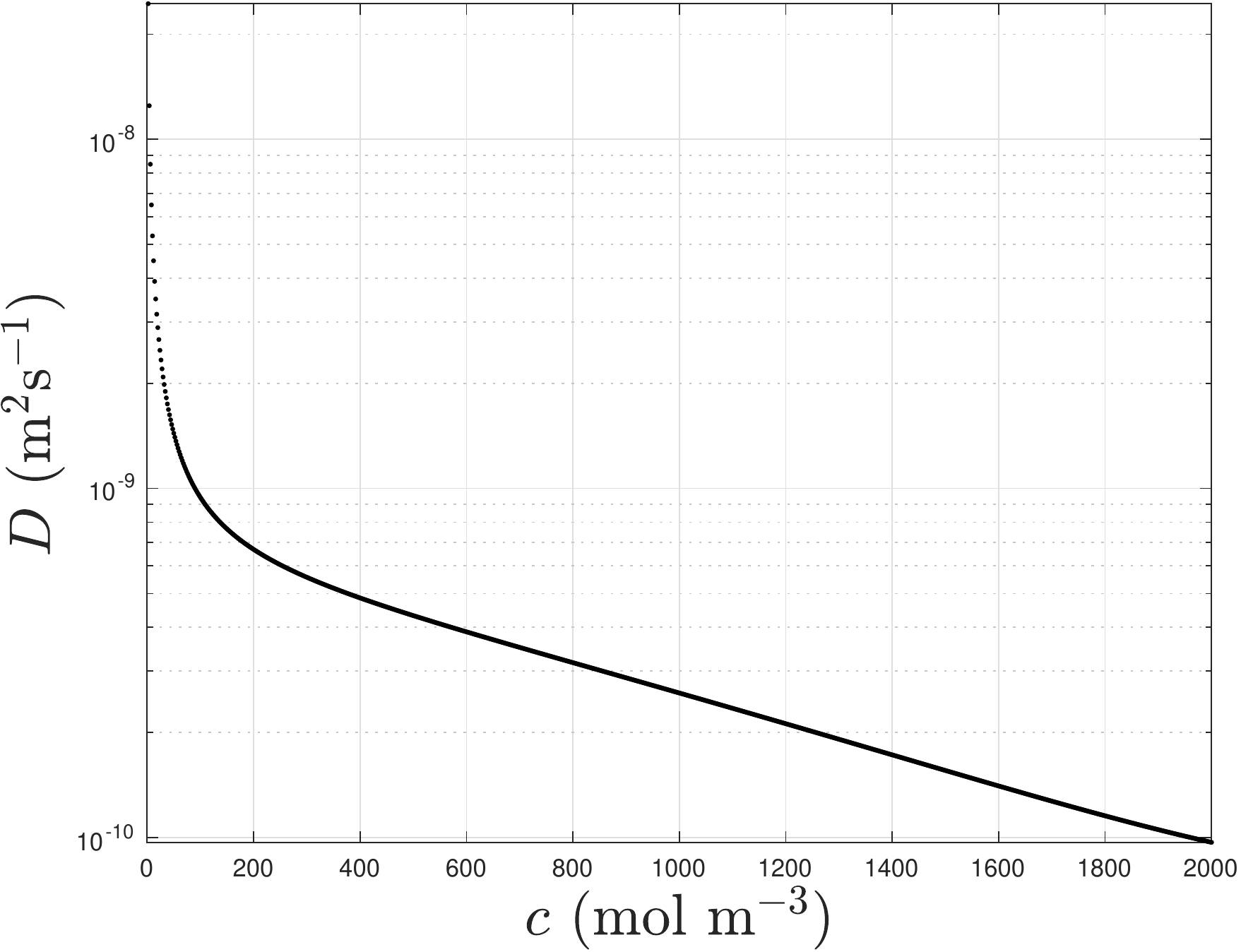}}\quad
    \subfigure[]{\includegraphics[width=7cm]{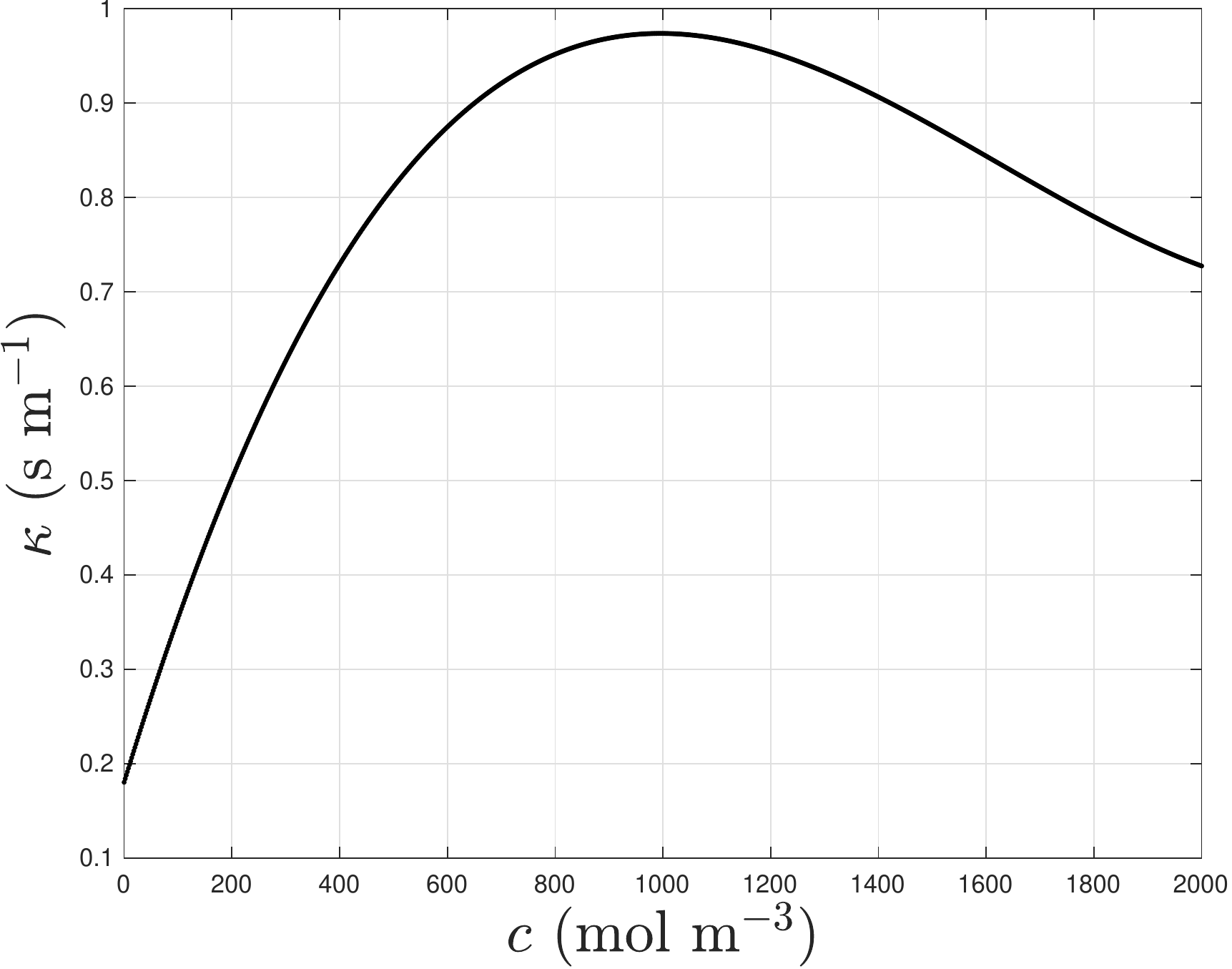}}}
  \mbox{
    \subfigure[]{\includegraphics[width=7cm]{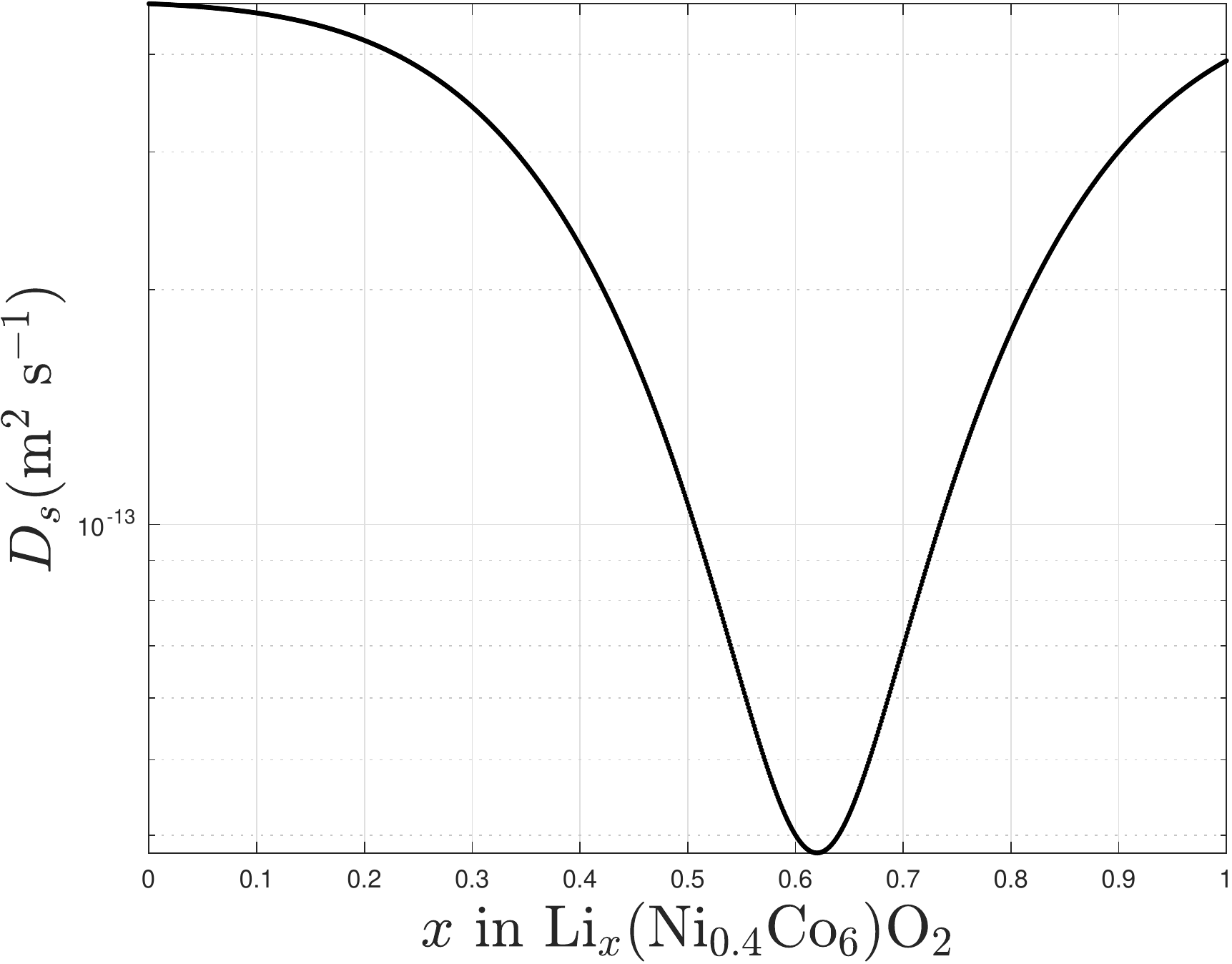}}\quad
    \subfigure[]{\includegraphics[width=7cm]{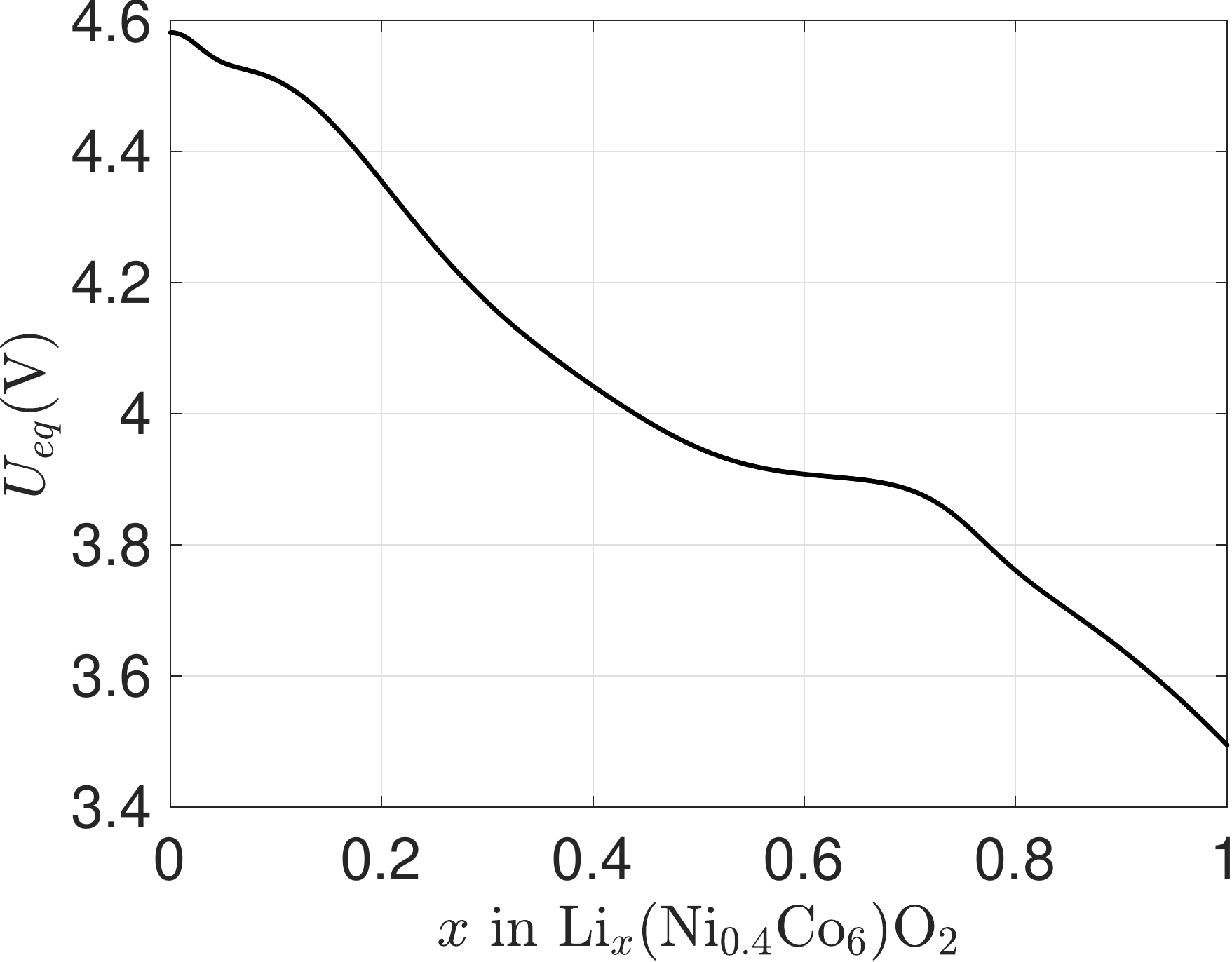}}}
  \caption{{State-dependent material properties: (a) ionic
      diffusivity, (b) conductivity of {the liquid phase}, (c)
      diffusivity in the solid {phase} and (d) equilibrium potential
      of Li(Ni$_{0.4}$Co$_{0.6}$)O$_2$.}}\label{fig:params}
\end{figure}

Our aim is to assess the uncertainty in the calibration of the SP and
cSP models by solving suitably defined Bayesian inverse problems where
the ``measurements'' will be generated by solving the DFN model
{and will serve} as psuedo-experimental data. By framing the
problem in this way we eliminate various experimental measurement
errors and this will allow us to concentrate on sources of calibration
uncertainty intrinsic to the models themselves.  We focus on modelling
a half-cell made from nickel manganese cobalt oxide
Li(Ni$_{0.4}$Co$_{0.6}$)O$_2$ and shown schematically in Figure
\ref{fig:halfcell}. A complete data set of the material properties and
constitutive relations characterizing this cell for the purposes of
solving the DFN model is collected in Table \ref{tab:known}. The
parameters that we will later take to be unknown for the purposes of
inverse modelling are marked with a dagger ($^\dag$). Exactly which
parameters can be determined by inverse modelling depends upon the
model being utilized; many parameters in the DFN model do not appear
in the simplified models. This information is summarized in Table
\ref{tab:unknown}. In the case of the cSP model we elect to take the
vector of unknown parameters to be $\bP = [ \B, \sigma_s, \widehat{D},
R, \widehat{D}_s ]$. Here the quantities $\widehat{D}$ and
$\widehat{D}_s$ might aptly be referred to as the ``characteristic
size'' of the diffusivity in the electrolyte and LNC respectively.
More precisely they are defined as follows
{\begin{align}
    D(x) &= \widehat{D} \, [0.278x^3 - 1.356x^2 + 1.87x + 0.180]x^{-1},\\
    D_s(x) &= \widehat{D}_s \, [4.7 - 4.3\exp(-12(x - 0.62)^2)],
\end{align}
cf.~Figures \ref{fig:params}(a,c),} so that changing the hatted
quantities corresponds to maintaining the functional form of the
diffusivity but {modifies its magnitude} by a multiplicative
factor. We frame the inverse modelling problem using the hatted
quantities (rather than the functions {$D(x)$ and $D_s(x)$}
themselves) because it is markedly simpler to infer a scalar quantity
than a function.  {This being said, the latter task has been
  accomplished in \cite{skhgp15a,ekfkgp20a}.}  In the case of the SP
model the situation is more nuanced since the parameters $R$ and
$\widehat{D}_s$ enter as a product and therefore cannot be
independently determined.  Thus, for the SP model we will assume that
the particle radius is known $R=6.5 \mu m$, cf.~Table \ref{tab:known},
and will infer the diffusivity coefficient $\widehat{D}_s$ such that
in this case the vector of unknown parameters will be $\bP = [
\widehat{D}_s ]$.

We will consider the problem of discharging the cell under different
rates, namely, 2C, 4C, 8C and 16C. This will allow us to see how the
discharging rate affects the uncertainty of parameter estimation.
Since time-dependent discharge voltage curves are relatively easy to
obtain in experiments, we will use these quantities predicted the DFN
model at the different discharging rates as our ``measurements''. They
will be denoted $\widetilde{V}(t)$ for $0 \le t \le t_f$, where $t_f$
is the duration of the simulated experiment which depends on the
C-rate. More specifically, for each C-rate the discharge time $t_f$ is
chosen {to achieve the same voltage drop, namely, such that
  $\widetilde{V}(t_f) = 3.58$ V, see Figures \ref{fig:V}(a,c,e,g) further
  below. This ensures that close to 100\% of charge is extracted from
  the cell and gives $t_f =0.529, 0.254, 0.109, 0.037$} hrs for 2C,
4C, 8C and 16C.  To summarize, our goal will be to infer ``true''
values of the parameters $\bP$ using the SP and cSP models based on
``measurements'' generated by the more complete DFN model. The
specific question we want to address is how the accuracy and
uncertainty of this calibration process depends on the model and the
C-rate.

\begin{table}[H]
\centering
\begin{tabular}{|c|c|c|}
\hline \textbf{Param./}  & \textbf{Description}  & \textbf{Value and Units} \\
\textbf{Function}  &  &  \\
\hline $L$ & Electrode thickness & 54 $\mu$m\\
\hline $R^\dag$ & Electrode particle radius & 6.5 $\mu$m \\
\hline $A$ & Electrode cross-section area  & 8.585 $\times 10^{-3}$ m$^2$ \\
\hline $\epsilon_l$ & Volume fraction of electrolyte in electrode matrix & 0.296 (dim'less) \\
\hline $ b$ & Particle surface area  per unit volume  electrode &3.249 $\times 10^{5}$  (m$^{-1}$) \\
\hline $k$ & Reaction rate constant & 5.904$\times 10^{-11}$ (mol$^{-1/2}$m$^{5/2}$s$^{-1}$)\\
\hline $c_s^{max}$ & Maximum concentration of Li$^+$  in solid & 28176.4 (mol m$^{-3}$)\\
\hline $c_{\text{init}}$ & Typical concentration of Li in electrolyte & 1000 (mol $m^{-3}$)\\
\hline $\tp$ & Transference number & 0.26 (dim'less) \\
\hline $\cal B^\dag$ & Permeability factor in electrode matrix & 0.153 (dim'less) \\
\hline $\sigma_s^\dag$ &  Conductivity in solid & 68.1(Sm$^{-1}$) \\
\hline $\ueq$ & Equilibrium potential & Figure \ref{fig:params}(d) (V) \\
\hline  $I$  & Current flow into cell & 0.15625 (A)\\
\hline
\hline $\widehat{D}^\dag$ & Typical diffusivity in electrolyte & 2.549$\times 10^{-10}$    (m$^2$  s$^{-1}$) \\
\hline $\widehat{D}_s^\dag$ & Typical diffusivity in solid & $10^{-13}$ (m$^2$  s$^{-1}$) \\
\hline $\widehat\kappa$ & Typical conductivity in electrolyte & 1 ( sm$^2$  ) \\
\hline
\hline $T$ & Absolute temperature & 298.15 (K)  \\
\hline $F$ & Faraday's constant & $96487 $(A\,s\,mol$^{-1}$)  \\
\hline $R_g$ & Universal gas constant & $8.3145 $ (J\,K$^{-1}$mol$^{-1}$)  \\
\hline
\end{tabular}
\caption{Physical constants and material properties characterizing a 
  cathode made from  Li(Ni$_{0.4}$Co$_{0.6}$)O$_2$. The material properties 
  which are not constant and are functions of state parameters are shown in 
  Figure \ref{fig:params}. Quantities denoted with a {dagger  ($^\dag$)} are
  considered unknown and  will be inferred via inverse modelling, cf.~Table \ref{tab:unknown}.}
\label{tab:known}
\end{table}
\begin{table}[H]
	\centering
	\begin{tabular}{|p{2.1cm}|p{2.5cm}|p{2.5cm}|p{2cm}|  }
		\hline
		Parameters& DFN  Model & cSP model &SP model\\
		\hline
		$\cal B$ & $\checkmark$   & $\checkmark$&   \\
		$ \sigma_s $ & $\checkmark$  &  $\checkmark$&  \\
		$ \widehat{D}$  & $\checkmark$  &  $\checkmark$&  \\
		$R $ & $\checkmark$    &  $\checkmark$& $\checkmark$\\
		$\widehat{ D}_s$& $\checkmark$    &  $\checkmark$   & $\checkmark$\\
		\hline
	\end{tabular}
	\caption{Unknown material properties which can be inferred via 
		inverse modelling  using different models (a tick $\checkmark$ indicates 
		that the given parameter appears in the model).}  
	\label{tab:unknown}
\end{table}

\section{Inverse Modelling and Uncertainty Quantification}
\label{sec:inverse}

Unknown model parameters can be inferred by choosing them so as to
minimize a least-square error between the corresponding model
predictions and the measurements. This error can be expressed as the following
functional
\begin{equation}
\mathcal{J}(\bP)= \int_{0}^{t_f} | V(t;\bP) - \widetilde{V}(t)|^2\, dt,
\label{eq:J}
\end{equation}
where $V(t;\bP)$, $0 \le t \le t_f$, is the voltage predicted by the
SP or cSP model with parameters given in $\bP$ and $t_f$ is the
duration of the simulated experiment which depends on the C-rate as
explained in the previous section.  Optimal values of these parameters
can in principle be determined by solving the optimization problem
\begin{equation}
\min_{\bP \in \mathbb{R}^d} \mathcal{J}(\bP).
\label{eq:min}
\end{equation}
This problem is trivial to solve for the SP model where $d=1$. One can
also solve it for the cSP model where $d=5$, however, as will be shown
in Section \ref{sec:results}, model parameters inferred in this way
are characterized by a very high level of uncertainty.

In order to quantify the uncertainty arising in estimation due to
model-reduction errors and {the fact that inverse problems are
  usually severely underdetermined}, we use a state-of-the-art
technique based on Bayesian inference \cite{10.5555/2568154}.  In this
approach {a} probabilistic setting is adopted as a way to quantify
uncertainty resulting from incomplete knowledge about the problem.
When our measurements are incomplete and/or noisy and our model is
inaccurate, then an inverse problem will admit many approximate
solutions.  Bayesian inference is an elegant and consistent framework
allowing one to use a combination of prior knowledge and experimental
data in order to assign specific confidence to different
reconstructions of the material properties.  Therefore, here we will
represent the reconstructed material properties $\bP$ in terms of
random variables characterized by certain probability density
functions (PDFs).

In the Bayesian framework the probability distribution of the inferred
material properties is given in terms of the {\it posterior}
probability $\pi(\bP | \tV)$ that the material properties take values
$\bP$ given the entire set of observations $\tV$. This can be
expressed using Bayes' Theorem, $\pi(A|B) \pi(B)=\pi(B|A) \pi(A)$,
where $A$ and $B$ are events \cite{s10,10.5555/2568154,Tenorio}, in
the form
\begin{equation}
  \pi(\bP|\tV)=\frac{\pi(\tV | \bP) \, \pi_0(\bP) }{\pi(\tV)}.
\label{eq:bayes}
\end{equation}
Here $\pi_0(\bP)$ is the {\it prior} distribution reflecting our {\it
  a priori} assumptions about the solution; in practice it may be
based, for example, on previous studies in the literature estimating
the material properties in question.  The term $\pi(\tV| \bP)$ is the
likelihood of observing particular experimental data $\tV$ for a given
set of material properties $\bP$, while $\pi(\tV)$ is the overall
probability of observing the experimental data $\tV$ and can be
treated as a normalizing factor.

In the present study we will adopt a {piecewise-constant prior
  $\pi_0(\bP)$ which is nonzero for values of each parameter $\bP$ in
  a large interval where they can be generally considered ``physically
  acceptable'', and zero otherwise. The rationale for this approach is
  to rule out parameter value which are physically inadmissible (e.g.,
  because of a wrong sign or if they are off by a few orders of
  magnitude). Thus, for such physically acceptable parameter values
  $\bP$,} the posterior probability will be proportional to the
likelihood function.  As regards the likelihood function, the
following ansatz is typically adopted in Bayesian inference
\cite{10.5555/2568154,Tenorio,s10}
\begin{equation}
\pi(\tV|\bP)= \frac{ \exp\left(-{\mathcal{J}(\bP)}/{(2\sigma^2)}\right)}{\sigma\sqrt{2\pi}},
\label{eq:likfun}
\end{equation}
where $\sigma=0.005$ is the ``width'' selected such that if the value
of the error functional $\mathcal{J}(\bP)$ is below the numerical
accuracy with which the cell voltage is evaluated in the cSP model
(estimated at $\mathcal{O}(10^{-3})$ in Appendix \ref{sec:numerics}),
then $\bP$ is in a region where the likelihood distribution
concentrates 99.7\% of its probability
\cite{GnedenkoB.V.BorisVladimirovich1961Aeit}. This choice is made in
order to avoid overfitting (i.e., to avoid fitting to noise due to
errors incurred during spatial discretization and subsequent temporal
integration of the governing PDEs).  Expression \eqref{eq:likfun}
reflects the assumption that for a given set of material properties
$\bP$, measurements resulting in large values of the {error}
functional are less likely to be observed.  An intuitive motivation
for the choice of an exponential function in \eqref{eq:likfun} is that
in the hypothetical simplified case when the predicted discharge
voltage curves have a linear dependence on the material properties,
resulting in $\mathcal{J}(\bP)$ being a quadratic function of the
material properties $\bP$, relation \eqref{eq:likfun} would produce a
normal distribution which in the light of the central limit theorem is
universal.  A more rigorous justification of this choice can be found
for example in \cite{s10}.  The likelihood function $\pi(\tV |\bP)$ is
approximated by sampling the distribution in \eqref{eq:likfun} using
the Metropolis-Hastings algorithm \cite{chib1995understanding} to
produce $K$ samples {of the parameter vector $\bP$}.  The
normalizing factor $\pi(\tV)$ in \eqref{eq:bayes} is determined such
that the integral of the posterior probability $\pi(\bP | \tV)$ with
respect to the components of the vector $\bP$ is equal to unity.

The likelihood function \eqref{eq:likfun} is sampled using a suitably
designed random sequence of parameter vectors $\bP_i$, $i=1,\dots,K$,
the so-called Monte-Carlo Markov Chain (MCMC).  Given a sample
$\bP_i$, the next sample $\bP_{i+1}$ is generated using a proposal
function $\mathrm{Prop}(\bP_i \rightarrow \bP_{i+1})$ which in the
present study is taken in the form of a piecewise uniform distribution
proportional to the characteristic function of a small rectangular
neighborhood centered at $\bP_i$. If the new sample $\bP_{i+1}$ leads
to a decreased value of the likelihood function as compared to
$\bP_{i}$, then it is accepted with some probability; otherwise, it is
always accepted. This approach, referred to the Metropolis-Hastings
algorithm, ensures that while being generally attracted to regions of
the parameter space where the likelihood function attains large
values, samples belonging to the Markov chain are also allowed to
explore other regions of the parameter space.  It is known that as the
number of samples in the chain increases ($K \rightarrow \infty$),
this procedure produces an increasingly accurate approximation of the
posterior probability density $\pi(\bP|\tV)$. If the first sample
$\bP_i$ is not chosen well, the approach may require a ``burn-in''
period before elements of the chain reach regions of the parameter
space which are worth exploring. In order to minimize the effect of
the burn-in period, these initial samples are often removed from the
chain.  Our implementation of the MCMC Metropolis-Hastings approach is
described in Algorithm \ref{MHMCMC}.  In the computational results
reported in Section \ref{sec:results} we initialize the Markov chains
with the vector of true parameters $\bP_1 = \bP^* = [ \B^*,
\sigma_s^*, \widehat{D}^*, R^*, \widehat{D}_s^* ]$. The computed
Markov chains typically consist of $K = 10^4$ samples and we verified
that increasing this number does not significantly affect the results.
More information on Bayesian inference methods can be found in
\cite{Metropolis1953,alma991029576739705251}.
 
 \begin{algorithm}
   \caption{MCMC Metropolis-Hastings algorithm for the Bayesian
     approach to the solution of the inverse problem \eqref{eq:J}--\eqref{eq:min}.
 \newline {\bf Input:}  $\pi_0(\bP)$ --- prior distribution of parameters (uniform) \newline
 \hspace*{1.25cm} $\tV(t)$ --- measurements \newline 
 \hspace*{1.25cm} $\bP_0$ --- initial sample chosen such that $\pi_0(\bP_0) > 0$  \newline
 \hspace*{1.25cm} $K$ --- total number of samples in the chain \newline 
 \hspace*{1.25cm} $K_0$ --- length of the burn-in period ($0 \le K_0 \ll K$)  \newline 
 \hspace*{1.25cm} $\text{Prop}(\bP\rightarrow\bP')$ --- {proposal function} (uniform distribution) \newline
 \hspace*{1.25cm} $\rand(\text{Prop}(\bP\rightarrow\bP'))$
 --- random {sample of the proposal \newline \hspace*{5.75cm} (providing a new value of $\bP'$)}  \newline
  \hspace*{1.25cm} $\pi(\widetilde{V}(t)|\bP)$ --- likelihood of the measurements $\tV(t)$ given the parameter $\bP$, cf.~\eqref{eq:likfun}  \newline
 {\bf Output:}  $\{ \bP_i\}_{i=1+K_0}^K$ --- MCMC samples approximating the posterior distribution  $\pi(\bP|\tV)$  
 } 
\label{MHMCMC}
\begin{algorithmic}[1]
\Procedure{MCMC Metropolis-Hastings Algorithm}{}
\For {$i=1:K$}
\State $\bP_{i+1} = \rand(\text{Prop}(\bP_i \rightarrow \bP_{i+1}))$.
\State $\alpha = \min \left\lbrace 1, \frac{\pi(\widetilde{V}(t)|\bP_{i+1})\pi_0(\bP_{i+1})\text{Prop}(\bP_{i+1}\rightarrow \bP_i)}{\pi(\widetilde{V}(t)|\bP_i)\pi_0(\bP_i)\text{Prop}(\bP_i\rightarrow \bP_{i+1})} \right\rbrace \,$.
\State $\chi = \rand(0,1)$
\If {$\chi > \alpha$} 
\State {$\bP_{i+1}=\bP_i$}
\EndIf
\State \textbf{end};
\EndFor \textbf{end};
\State Remove the burn-in period if necessary
\EndProcedure
\end{algorithmic}
\end{algorithm}

\FloatBarrier

\section{Results}
\label{sec:results}

\begin{figure}
\centering 
\includegraphics[width=0.6\textwidth]{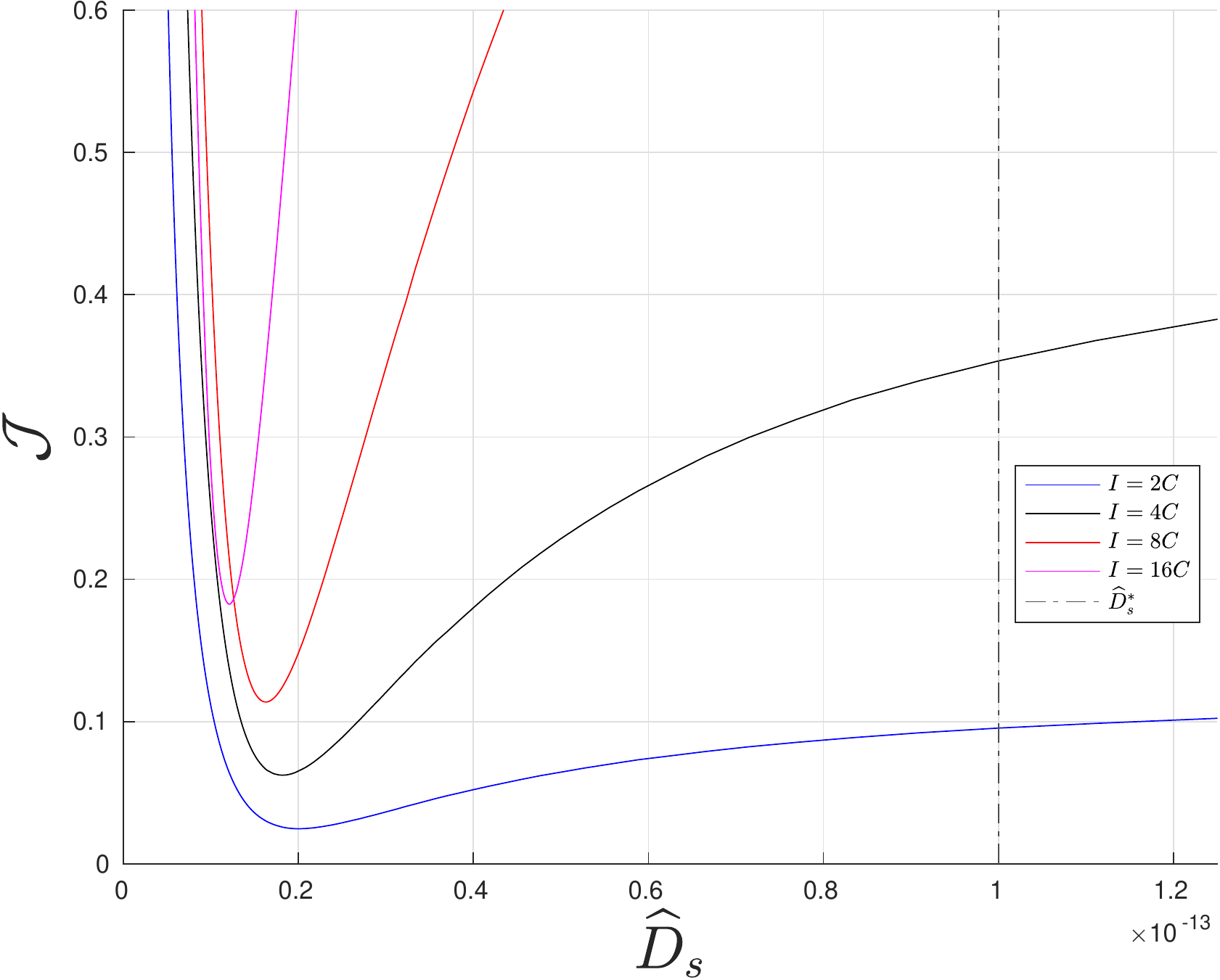}
\centering 
\caption{Dependence of the error functional $\mathcal{J}$ on the
  parameter $\widehat{D}_s$ in the SP model for the considered C-rates
  of 2C, 4C, 8C and 16C. The true value of the parameter
  $\widehat{D}^*_s=10^{-13}$ (m$^2$ s$^{-1}$) is indicated with a
  vertical line.}
\label{fig:SPM}
\end{figure}

In this section we present the solutions of the inverse problem
defined in \eqref{eq:J}--\eqref{eq:min} for the SP and cSP models,
beginning with the former.  In our framing of the problem, the SP
model depends on one unknown parameter only ($\widehat{D}_s$) and its
optimal value can be easily inferred by plotting the dependence of the
error functional $\mathcal{J}$ on $\widehat{D}_s$. These results are
shown for the different considered C-rates in Figure \ref{fig:SPM}
where we also indicate the true value $\widehat{D}^*_s$ that was used
in the DFN model to generate the measurements $\tV(t)$, cf.~Table
\ref{tab:known}. We remark that the values of the error functional are
in all cases much larger than the accuracy with which this quantity is
computed, estimated at $\mathcal{O}(10^{-3})$ in Appendix
\ref{sec:numerics}.

We note that the values of $\widehat{D}_s$ inferred based on the SP
model for different C-rates, corresponding to the minima of the error
functional $\mathcal{J}(\widehat{D}_s)$ in Figure \ref{fig:SPM}, do
not differ much between each other and that they all underestimate the
true value $\widehat{D}^*_s$ by a factor of more than 5. We also
observe that as the C-rate increases, the fits obtained with the SP
model become less accurate which is reflected in the minimum values of
the error functional becoming larger. At the same time, when the
C-rates are small, the minima of the error functional are shallow,
meaning that small values of the error functional, and hence also good
fits, are obtained for a broad range of values of $\widehat{D}_s$.
However, since shallow minima are harder to accurately capture in the
numerical solution of the optimization problem \eqref{eq:min}, this
property can be a source of a significant uncertainty of
reconstruction of $\widehat{D}_s$, especially in the presence of
experimental and numerical inaccuracies.

The consistent underestimation of $\widehat{D}_s$ produced by the SP
model can be rationalized as follows: The SP model is an approximation
of the DFN model where it is assumed that the only appreciable
potential drop is due to the equilibrium overpotential of the LNC
material, see \eqref{spm4}. Since it neglects the other potential
drops (i.e., those across the electrolyte, electrode and the
Butler-Volmer overpotential), good agreement between the SP and DFN
models is obtained by decreasing the value of $\widehat{D}_s$ in the
SP model which hinders transport of Li inside the LNC, boosting the
concentration of Li on the surface of the LNC particles and thereby
increasing the potential drop due to the equilibrium overpotential.
This compensates for the other drops that have been neglected and
brings the {voltage curves predicted by} the two models closer
together.

\begin{figure}
\centering 
\mbox{
\subfigure[]{\includegraphics[width=0.45\textwidth]{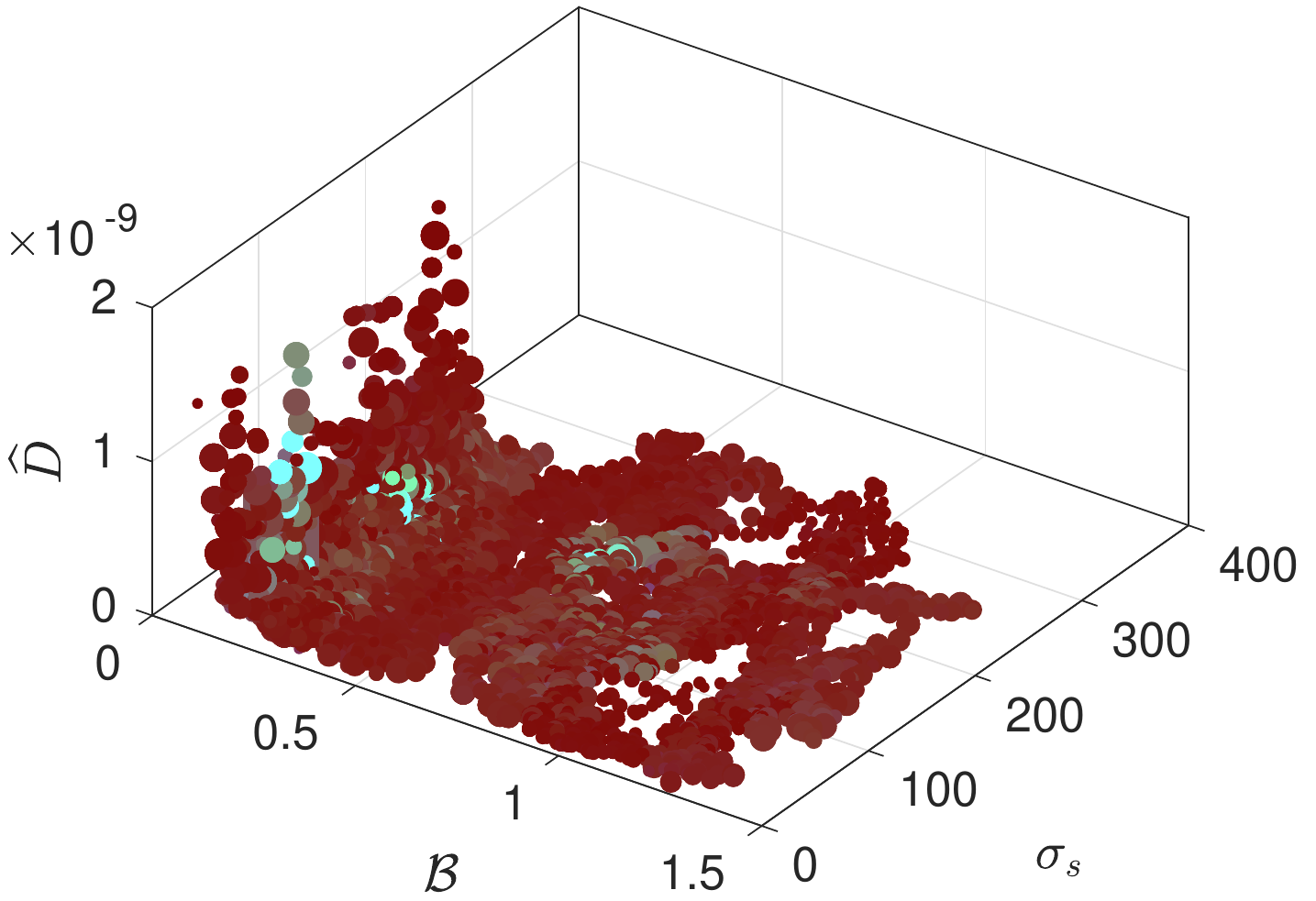}} \qquad
\subfigure[]{\includegraphics[width=0.45\textwidth]{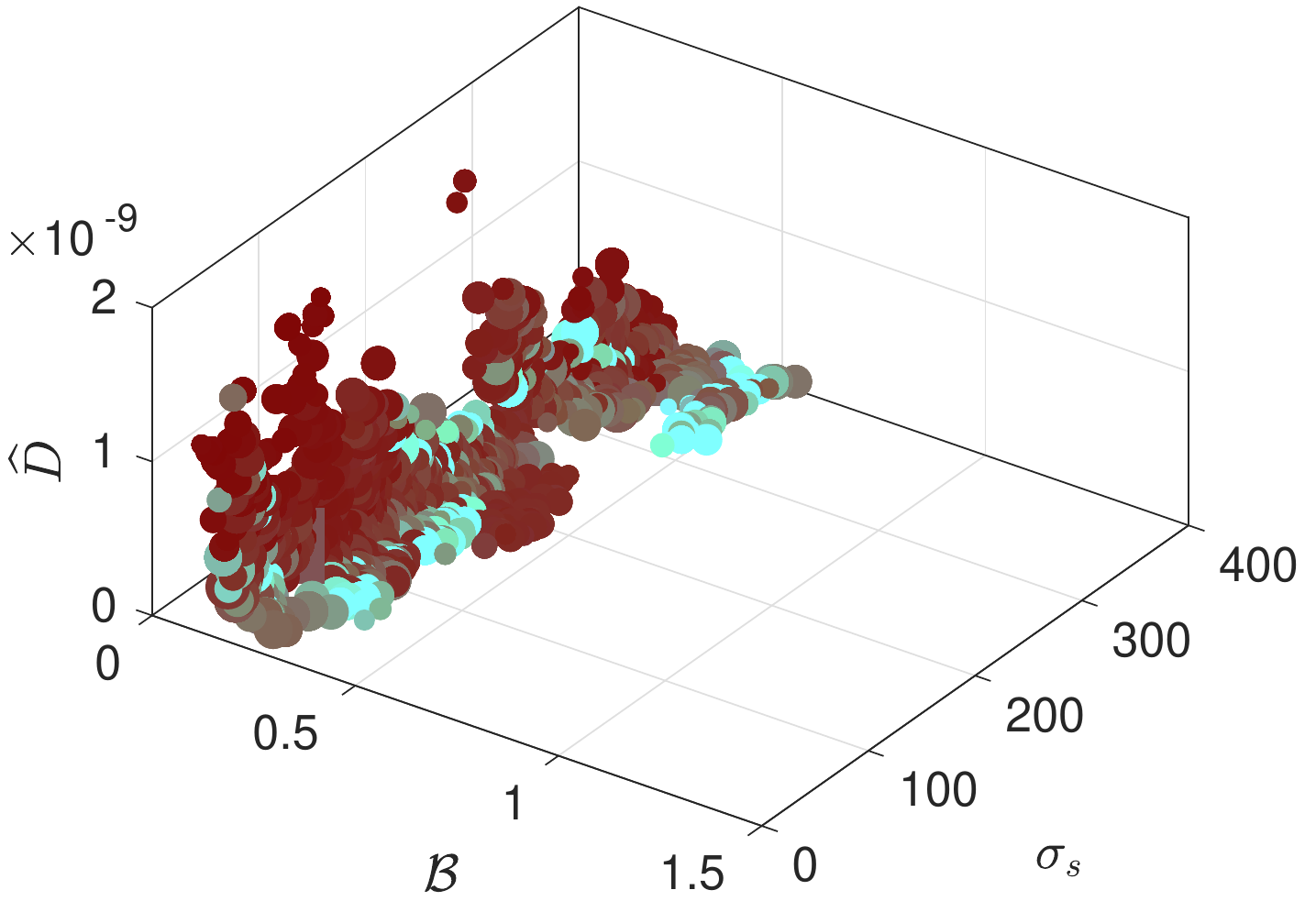}}}
\centering 
\mbox{
\subfigure[]{\includegraphics[width=0.45\textwidth]{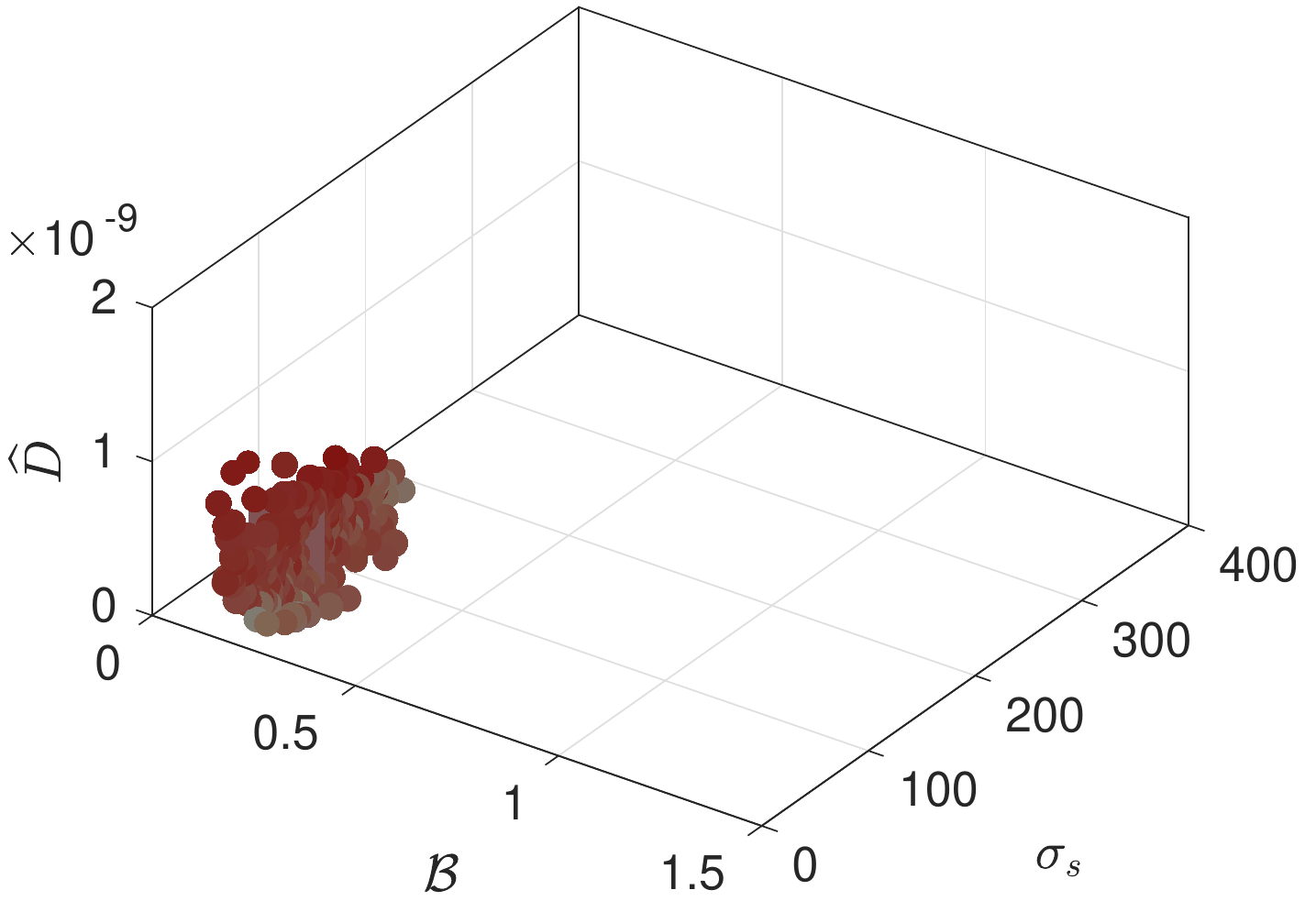}} \qquad
\subfigure[]{\includegraphics[width=0.45\textwidth]{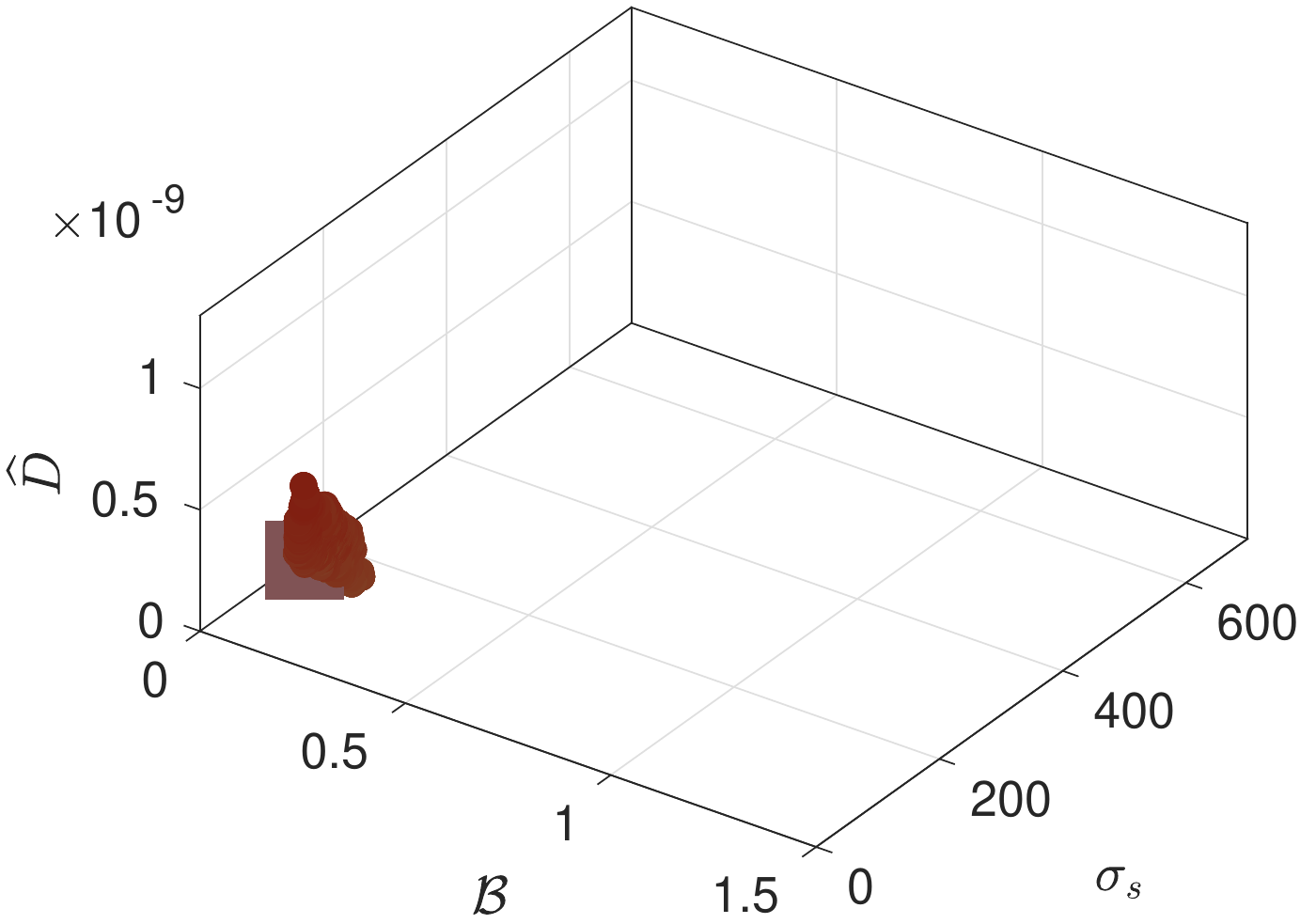}}}
\mbox{
\subfigure[]{\includegraphics[width=0.45\textwidth]{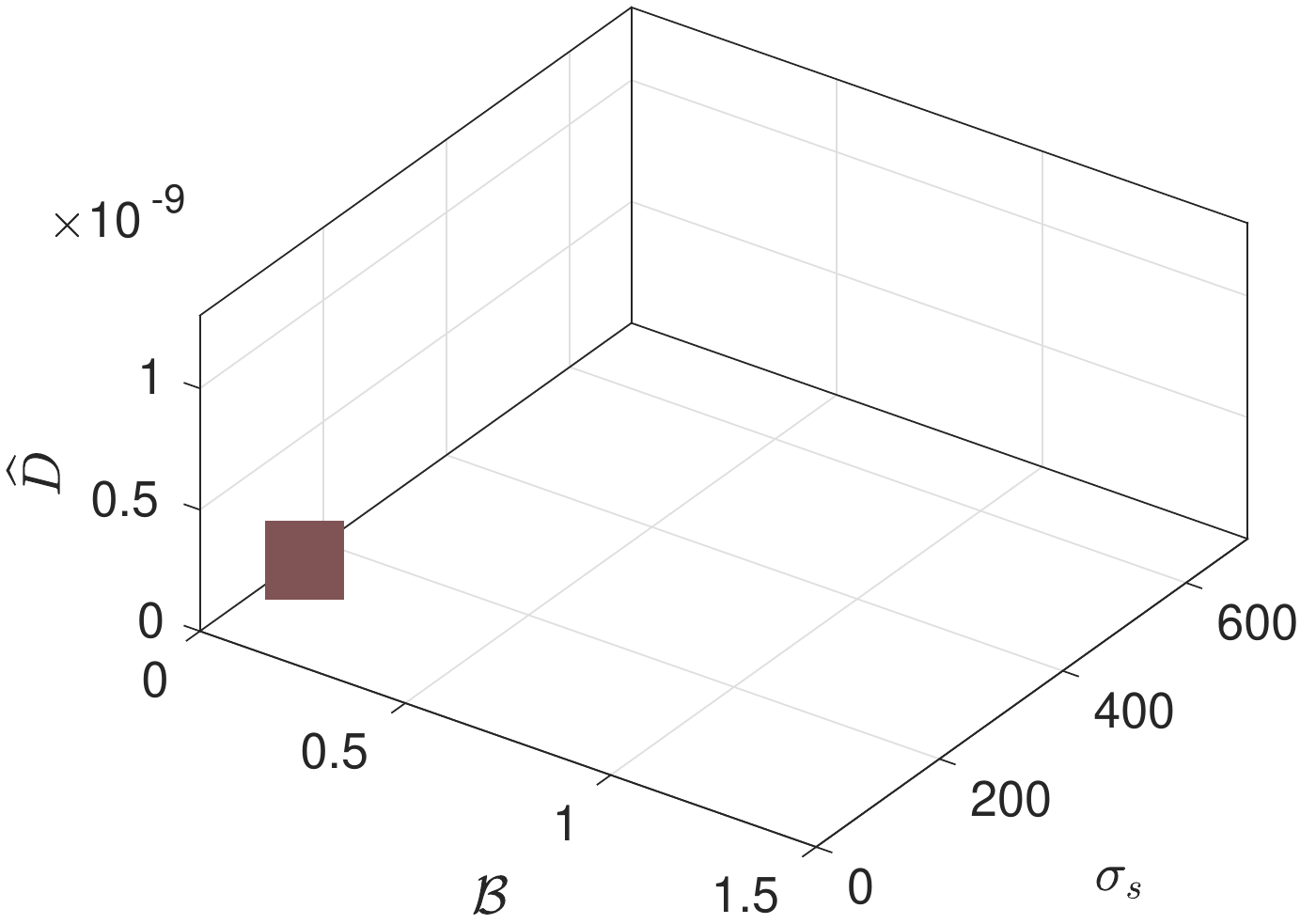}} \qquad
\subfigure[]{\includegraphics[width=0.45\textwidth]{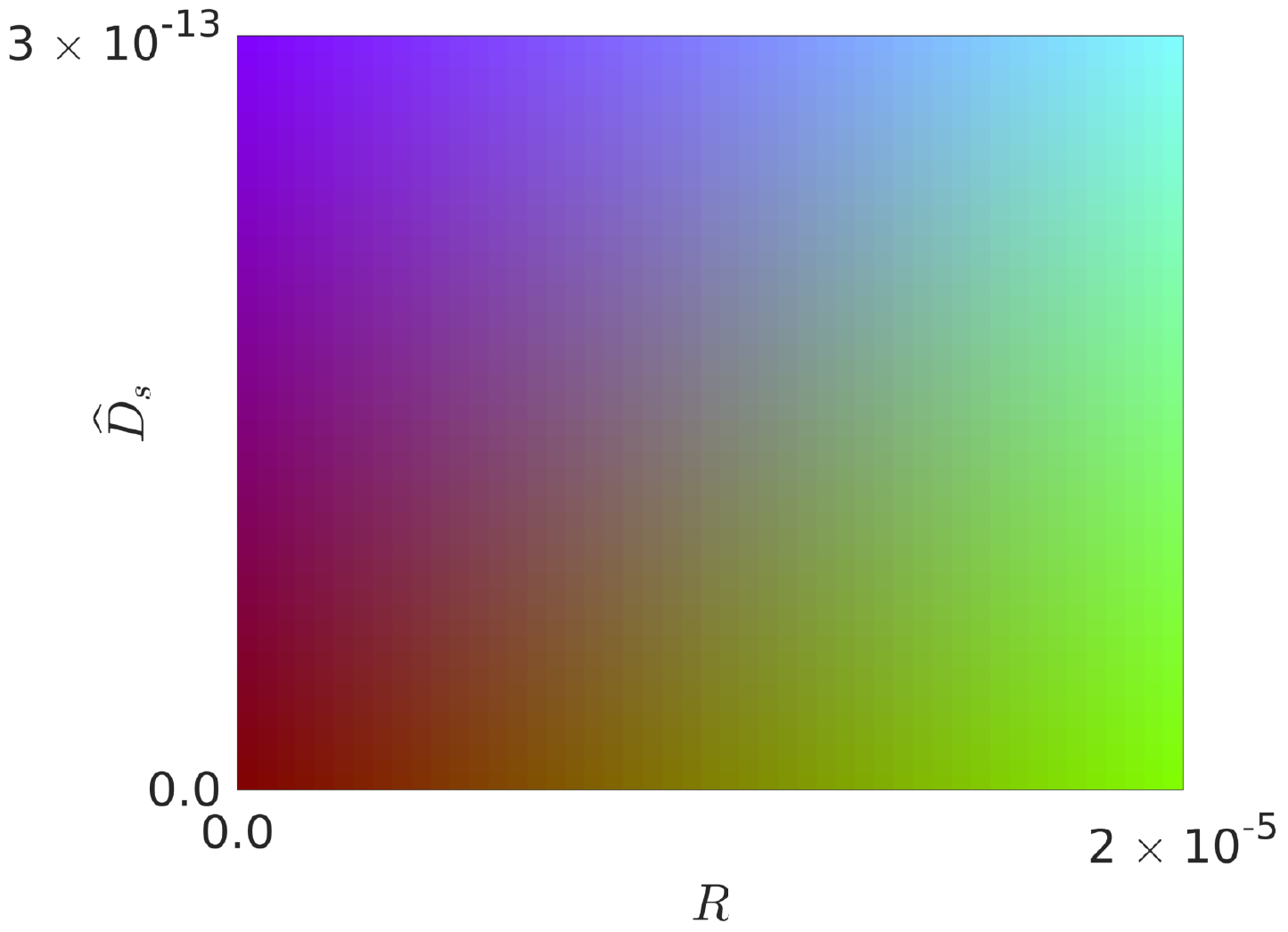}}}
\caption{Approximations of the the posterior probability distribution
  $\pi(\bP|\tV)$, cf.~\eqref{eq:bayes}, obtained in the Bayesian
  analysis of the inverse problem \eqref{eq:J}--\eqref{eq:min} for the
  C-rates of (a) 2C, (b) 4C, (c) 8C, (d) 16C. Symbols (circles)
  represents elements $\bP_i$, $i=1,\dots,K$, of the Markov chains
  with the parameters $\{ \B, \sigma_s, \widehat{D} \}$ represented in
  terms of the Cartesian coordinates and $\{R, \widehat{D}_s \}$
  encoded using the color scale given in panel (f).  The size of the
  symbols in panels (a)--(d) is proportional to
  $1/\mathcal{J}(\bP_i)^2$ (with different proportionality constant in
  each panel). The true parameter values are denoted with a big cube
  and are also shown separately in panel (e).}
\label{fig:RGB}
\end{figure}

\begin{figure}
\centering 
\mbox{
\subfigure[]{\includegraphics[width=0.45\textwidth]{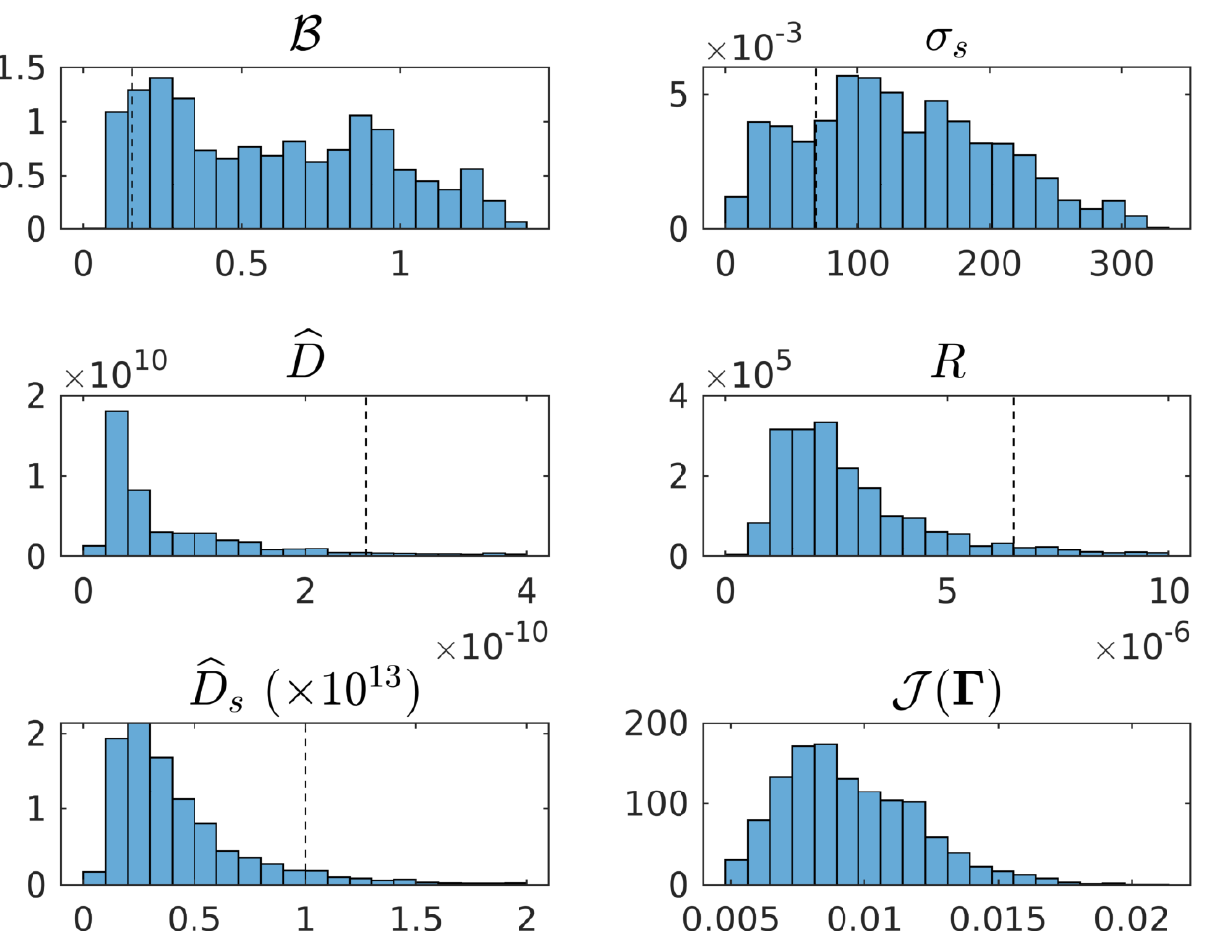}} \qquad
\subfigure[]{\includegraphics[width=0.45\textwidth]{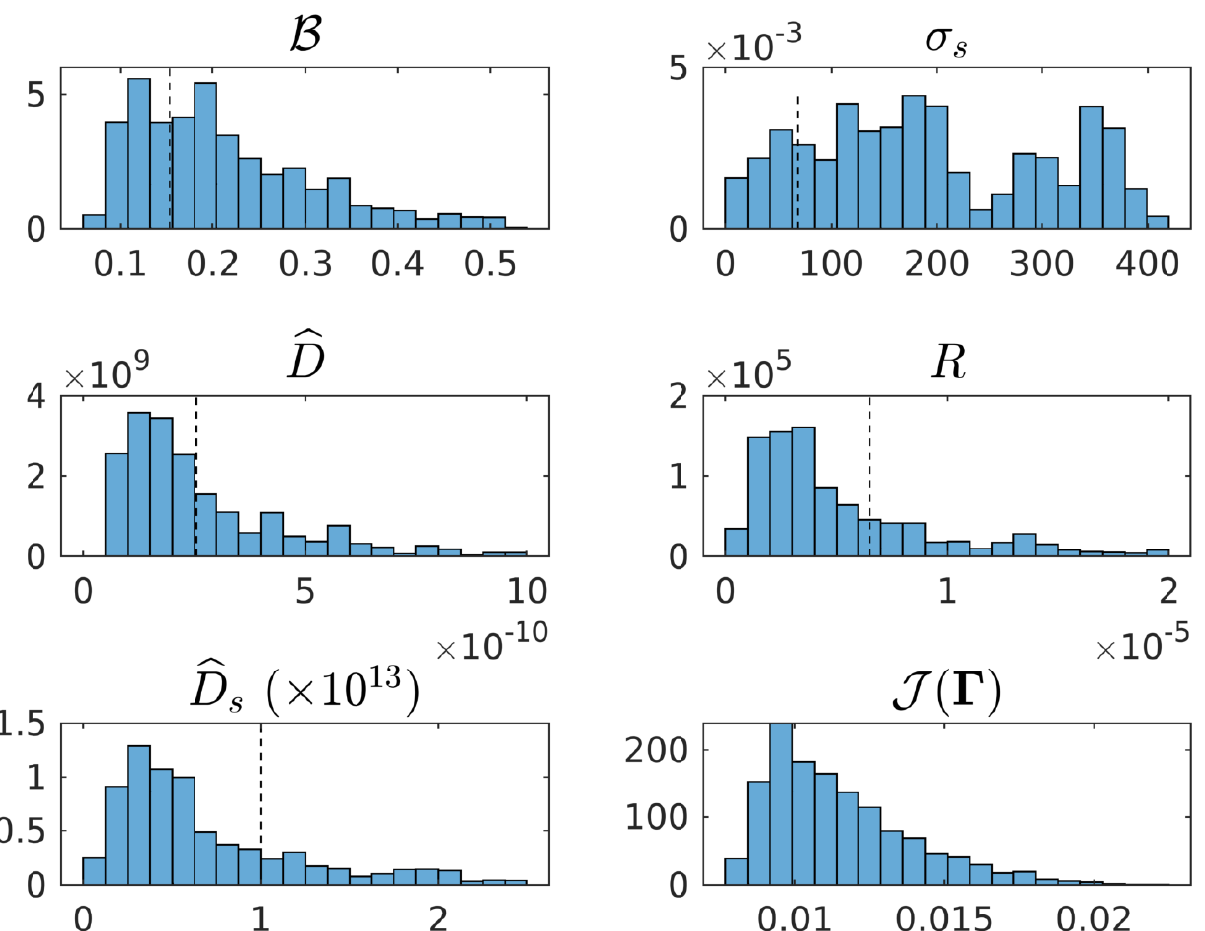}}}
\centering 
\mbox{
\subfigure[]{\includegraphics[width=0.45\textwidth]{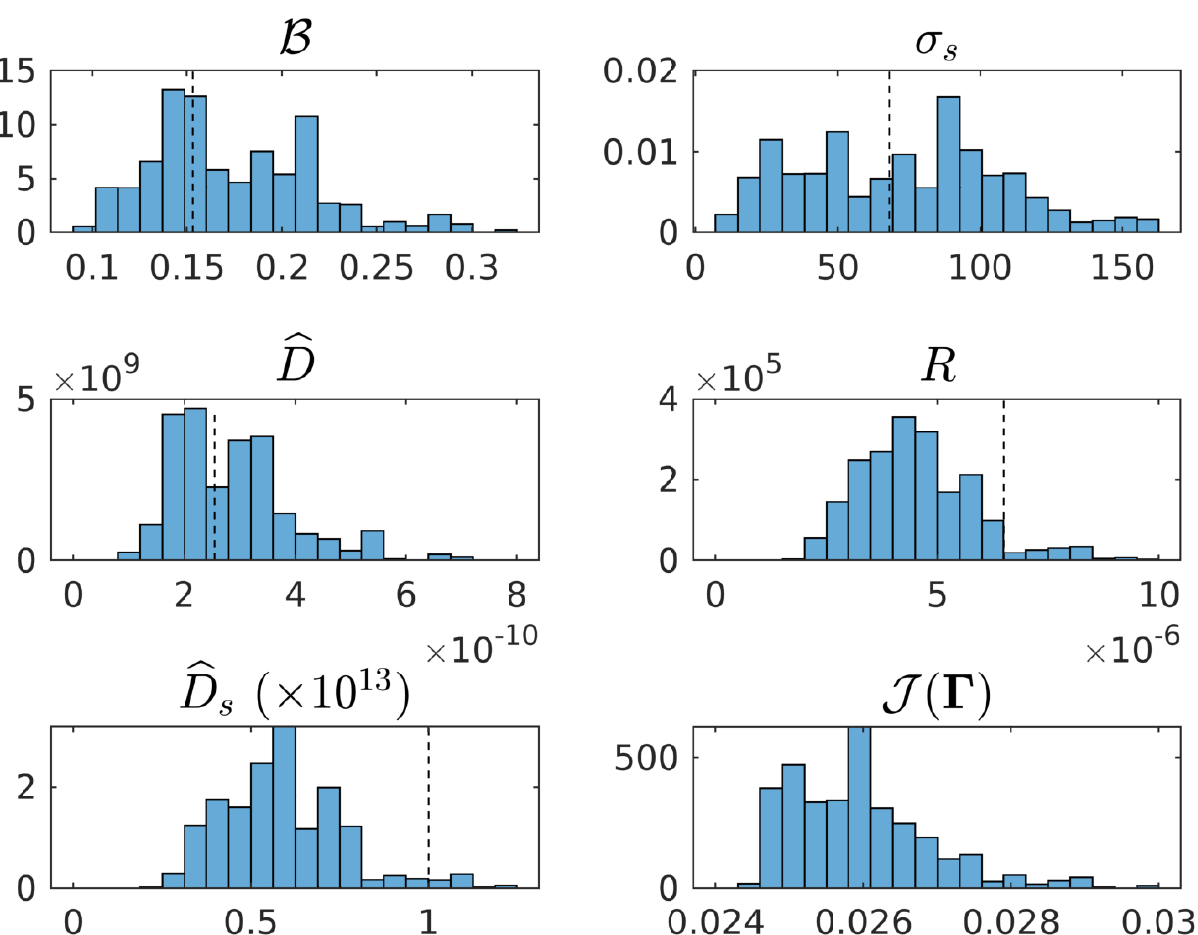}} \qquad
\subfigure[]{\includegraphics[width=0.45\textwidth]{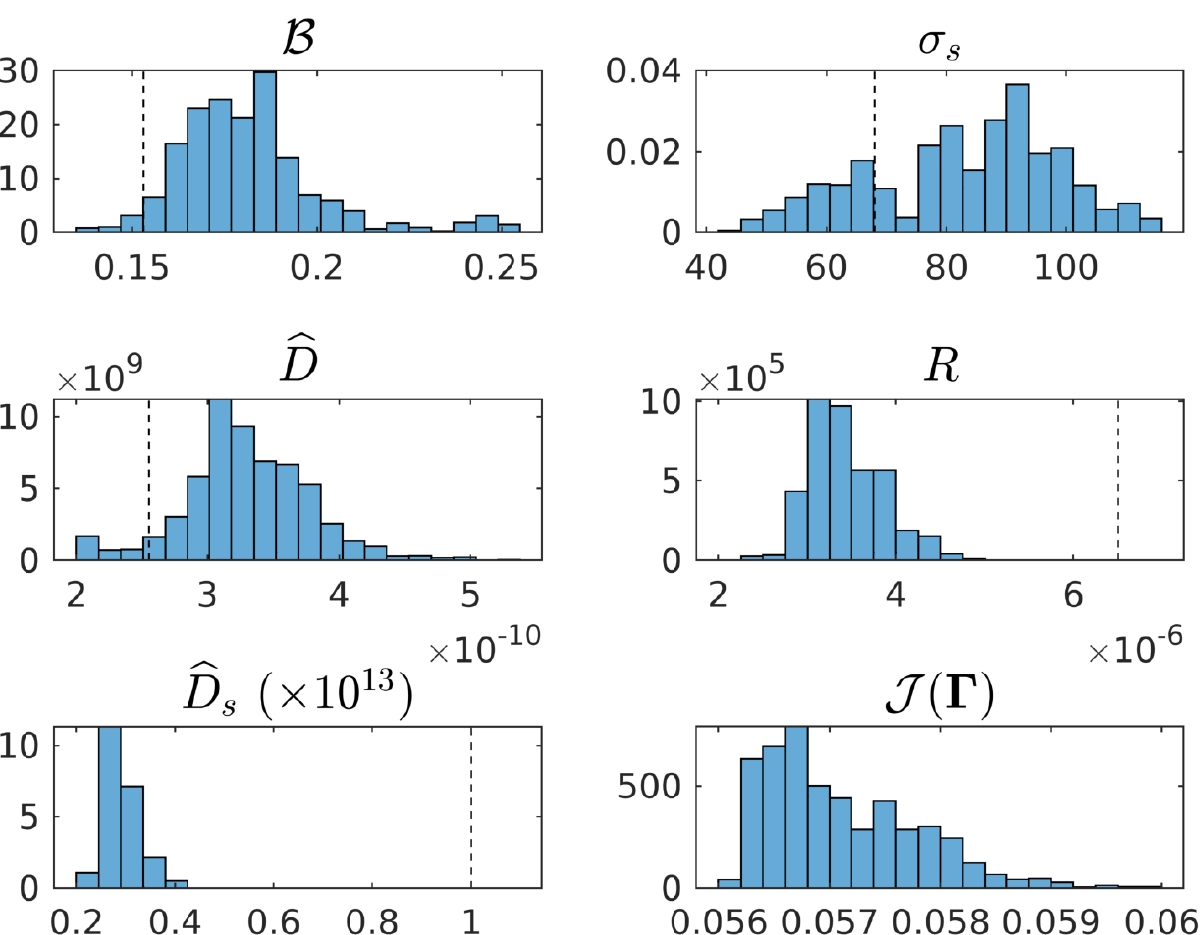}}}
\caption{Probability density functions of the material parameters $\bP
  =[ \B, \sigma_s, \widehat{D}, R, \widehat{D}_s ]$ and of the error
  functional $\J(\bP)$ obtained along the Markov chains for the
  different C-rates (a) 2C, (b) 4C, (c) 8C, (d) 16C.}
\label{fig:PDF}
\end{figure}

We now move on to present our solutions to the inverse problem
\eqref{eq:J}--\eqref{eq:min} for the cSP model, where the inferred
parameters are $\bP = [ \B, \sigma_s, \widehat{D}, R, \widehat{D}_s
]$. Because of the uncertainty inherent in these solutions, which will
become evident shortly, it is impractical to solve the inverse problem
directly by minimizing the error functional \eqref{eq:J} using methods
of numerical optimization. Instead, we adopt the Bayesian formulation
introduced in Section \ref{sec:inverse} where Algorithm \ref{MHMCMC}
is used to generate Markov chains consisting of a large number of
samples ($K = 10^4$). Each parameter sample $\bP_i$, $i=1,\dots,K$, is
associated with the corresponding value of the error functional
$\J(\bP_i)$. We recall that Algorithm \ref{MHMCMC} is designed to
explore regions of the parameter space characterized by large values
of the likelihood function \eqref{eq:likfun}. Samples in the Markov
chains obtained for the different considered C-rates are shown in
Figures \ref{fig:RGB}(a--d). Since there are five inferred parameters,
we visualize these results representing the parameters $\B$,
$\sigma_s$ and $\widehat{D}$ in terms of the three Cartesian
coordinates with the remaining two parameters $R$ and $\widehat{D}_s$
color-coded in terms of the red-green-blue (RGB) color scheme. The RGB
color model represents different colors by adding weighted
contributions, with weights varying between 0 and 1, of red, green and
blue.  In Figures \ref{fig:RGB}(a--e) the weight of red is chosen as
1/2, whereas the weights of green and blue are proportional to the
values of $R$ and $\widehat{D}_s$, as indicated in the color map shown
in Figure \ref{fig:RGB}(f) (due to this convention, red is the
dominant color in Figures \ref{fig:RGB}(a--e)). Moreover, the size of
the symbols (circles) representing a given sample $\bP_i$ is
proportional to $1/\J^2(\bP_i)$ such that parameter values producing
better fits are shown with larger symbols and are therefore more
visible (however, the proportionality constant is different for
different C-rates, such that symbols of the same size in Figures
\ref{fig:RGB}(a--d) do not correspond to the same values of the error
functional). Thus, the clouds of markers shown in these figures can be
interpreted as approximations of the posterior probability
distributions $\pi(\bP|\tV)$, cf.~\eqref{eq:bayes}.  In Figures
\ref{fig:RGB}(a--d) we also indicate the true values $\bP^*$ of the
parameters and for clarity they are also show separately in Figure
\ref{fig:RGB}(e). These results are complemented in Figure
\ref{fig:PDF} by the PDFs of the different material parameters $\bP =[
\B, \sigma_s, \widehat{D}, R, \widehat{D}_s ]$ and of the error
functional $\J(\bP)$ obtained along the Markov chains for the
different C-rates.

The key observation to be made about the results shown in Figures
\ref{fig:RGB} and \ref{fig:PDF} is that for decreasing C-rates the
parameter values producing good fits are increasingly scattered which
reflects the growing uncertainty of their reconstructions. For the
C-rate of 2C, cf.~Figure \ref{fig:RGB}(a), these parameters values
form a curved 2-dimensional ``manifold'' in the space spanned by
$\{\B, \sigma_s, \widehat{D}\}$ where the parameters values differing
by 300\% or more may still give rise to equally accurate fits.  For
4C, cf.~Figure \ref{fig:RGB}(b), the uncertainty of $\B$ is
significantly reduced, but at the price of increasing the
reconstruction uncertainty of the remaining parameters relative to the
case with 2C, which is evident from comparing the PDFs in Figures
\ref{fig:PDF}(a) and \ref{fig:PDF}(b).  On the other hand, for large
C-rates (8 and 16C) the obtained parameter values are clustered more
closely reflecting reduced uncertainty of the reconstruction. However,
for large C-rates the accuracy of the fits is also reduced as is
evident from the PDFs of the error functional shown in the bottom
right panels in Figures \ref{fig:PDF}(a--d). The optimal values of the
parameters reconstructed at large C-rates are further away from their
true values than at low C-rates (except for the parameter $\sigma_s$
which is reconstructed rather well at high C-rates). {This last
  observation is less evident from Figure \ref{fig:RGB}, but can be
  deduced from the PDFs shown in Figure \ref{fig:PDF}.}  Overall,
these observations are consistent with what we found in the case of
the SP model above.

At low C-rate the dominant contribution to the cell voltage is the
that arising from the equilibrium overpotential of the LNC (hence the
validity of the assumptions underpinning the SP and cSP models). Thus,
at low C-rates, alterations to the values of the parameters $\{ \B,
\sigma_s, \widehat{D} , R, \widehat{D}_s \}$ have a relatively small
effect on the cell voltage, and this is manifested in the large spread
of the clouds at low C-rates. Even though the cSP model better
recovers the true value of $\widehat{D}_s$ than the SP model, it still
consistently underestimates it. Moreover, at high C-rates, the size of
the underestimation increases with C-rate and we speculate that this
is for similar reasons to those described above for the SP model.  The
worsening of the match between the inferred and true values at 16C
compared to 8C can be attributed to the breaking down of the
assumptions underlying the cSP model (which require C-rates to not be
too high); at 16C we are entering a regime where the cSP model does
not accurately reproduce the discharge curves predicted by the DFN
model. This is borne out by the results presented in
\cite{Richardson20}.


\begin{figure}
\centering 
\mbox{
\subfigure[]{\includegraphics[width=0.35\textwidth]{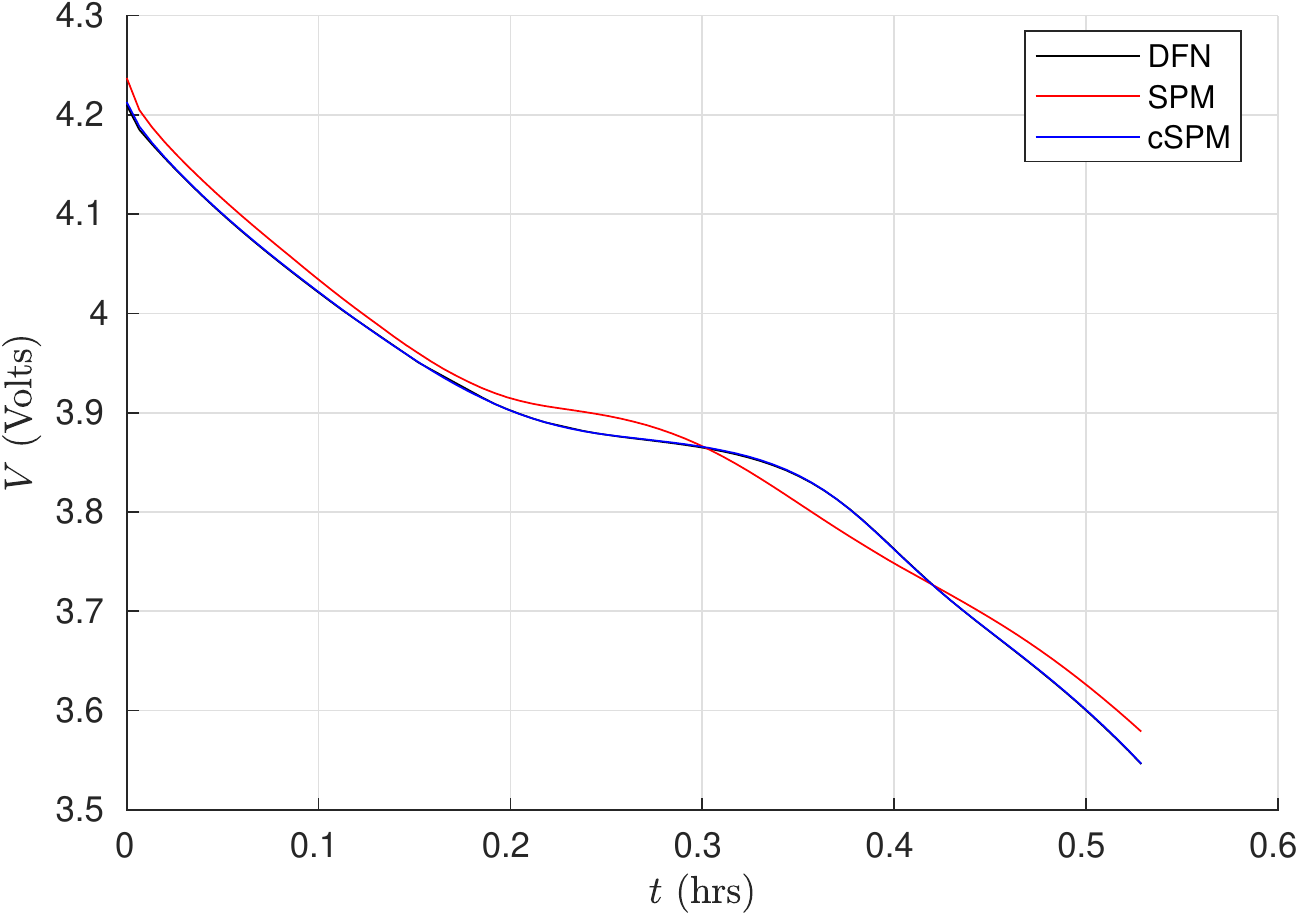}} \qquad
\subfigure[]{\includegraphics[width=0.35\textwidth]{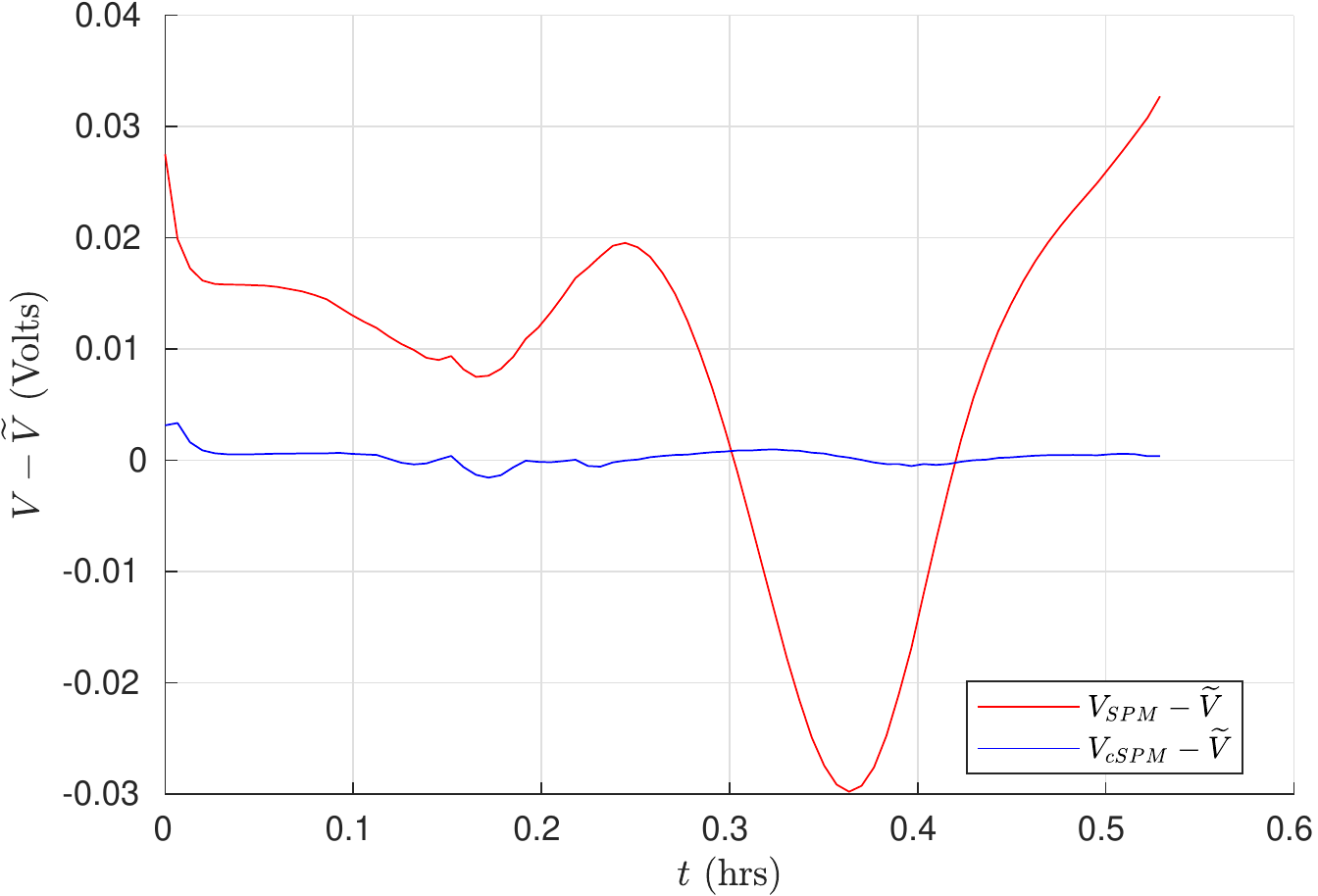}}}
\centering 
\mbox{
\subfigure[]{\includegraphics[width=0.35\textwidth]{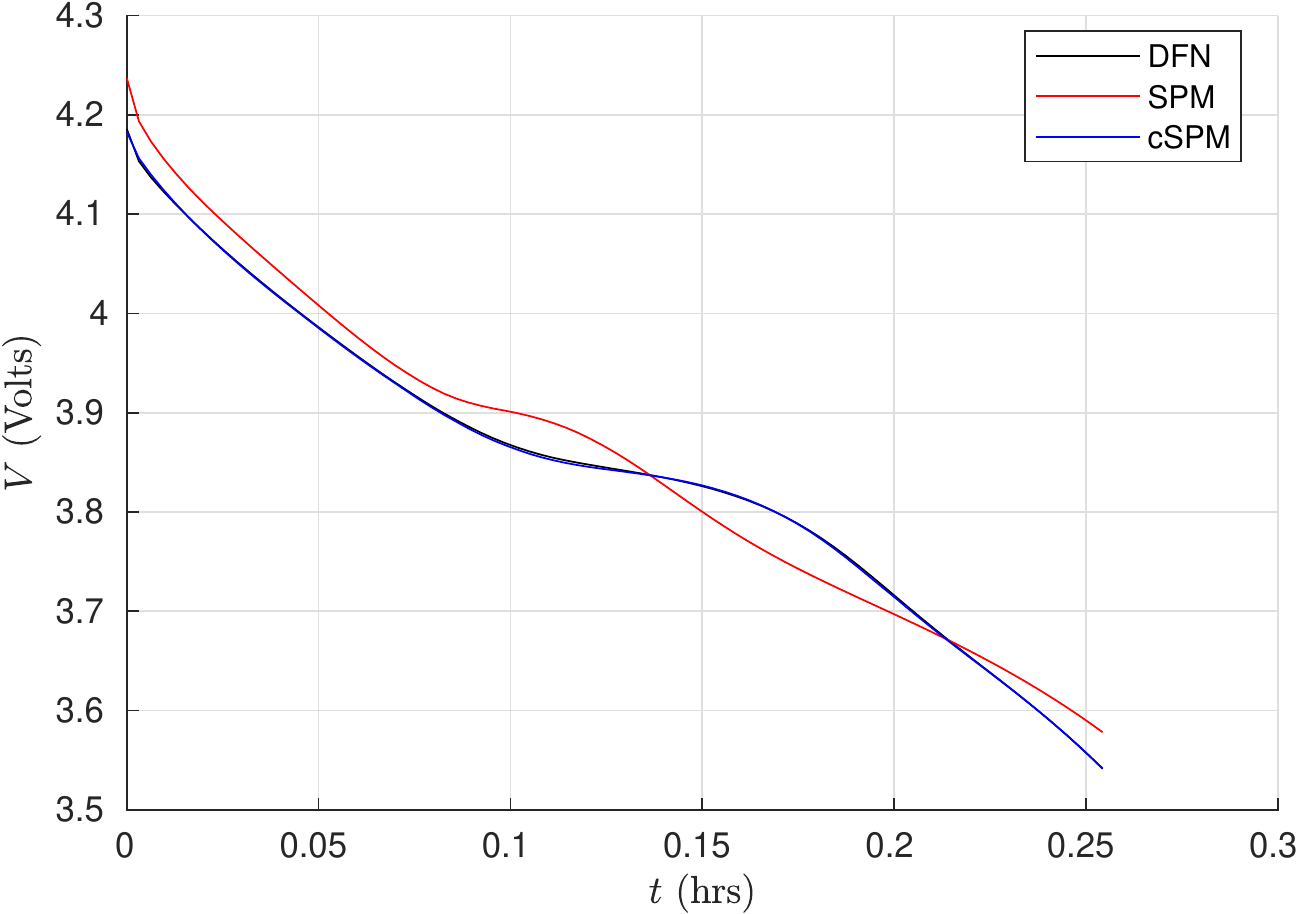}} \qquad
\subfigure[]{\includegraphics[width=0.35\textwidth]{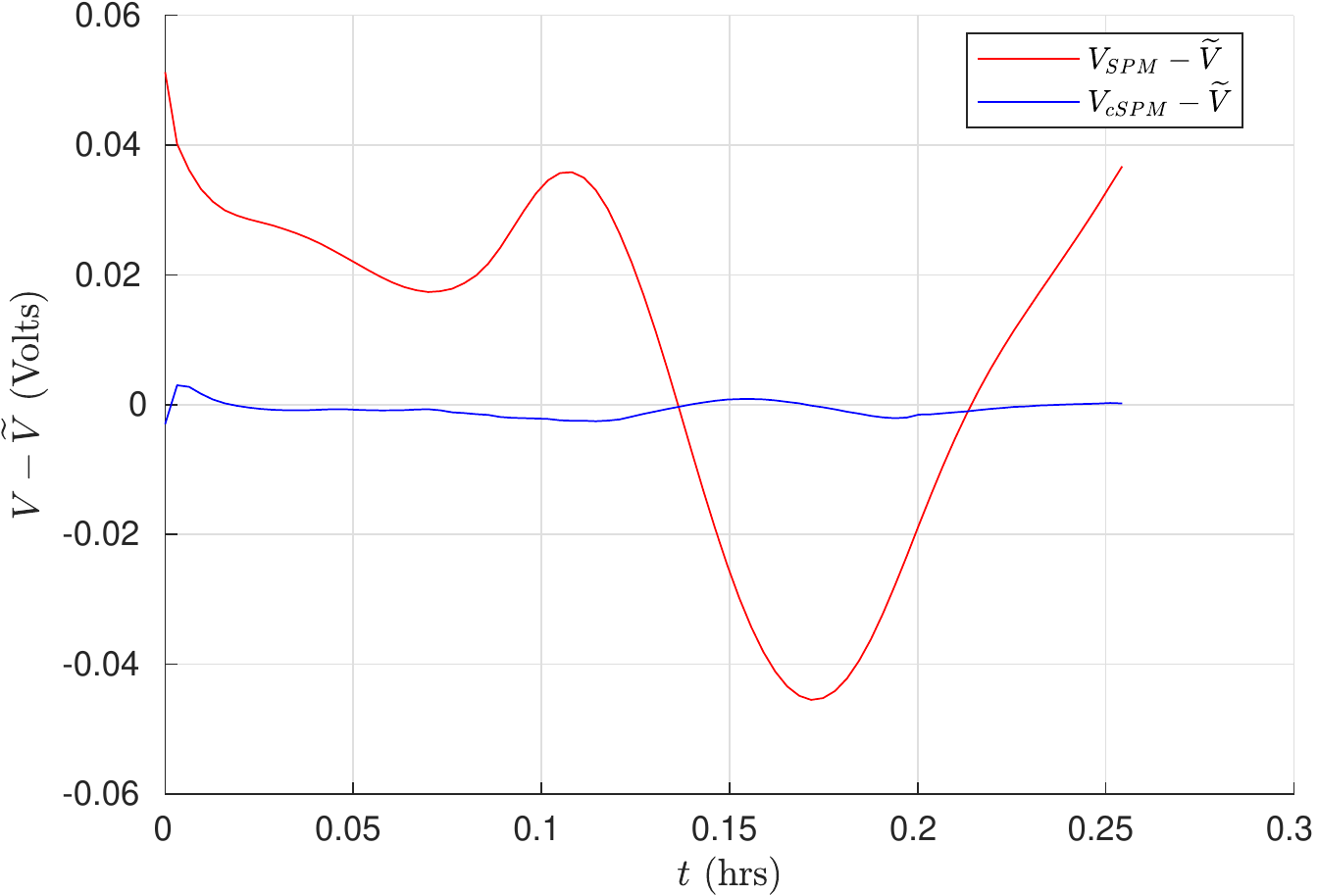}}}
\centering 
\mbox{
\subfigure[]{\includegraphics[width=0.35\textwidth]{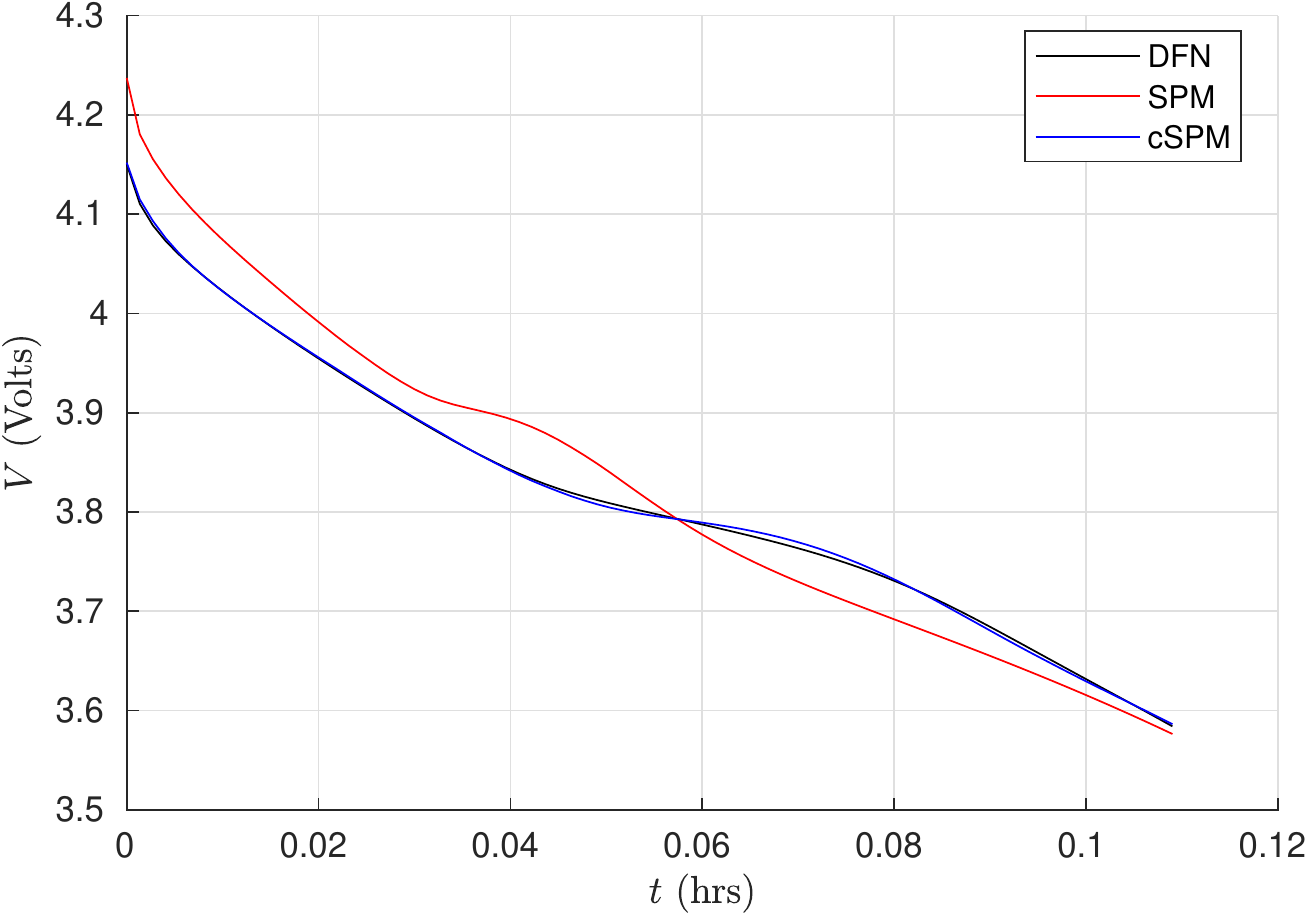}} \qquad
\subfigure[]{\includegraphics[width=0.35\textwidth]{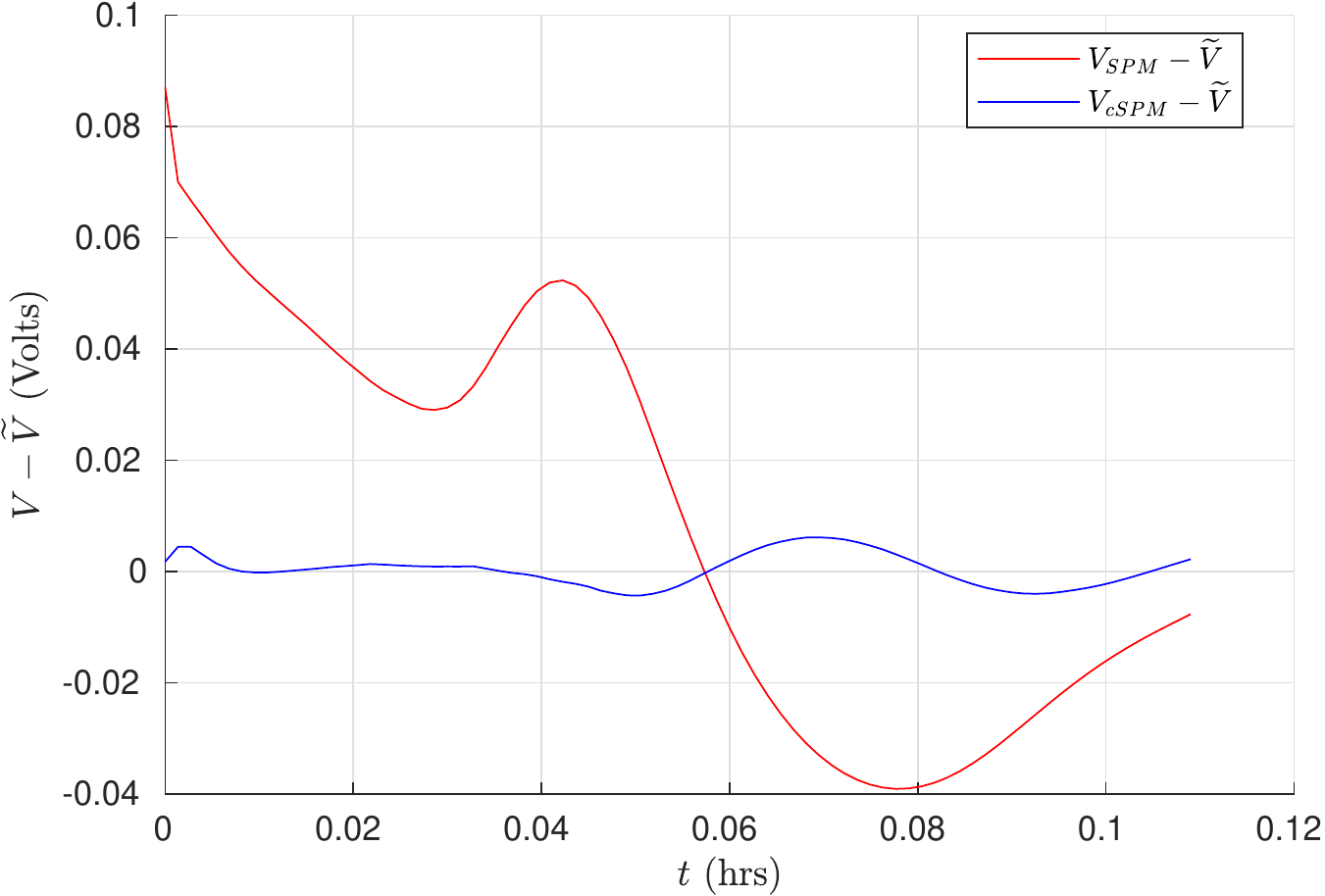}}}
\centering 
\mbox{
\subfigure[]{\includegraphics[width=0.35\textwidth]{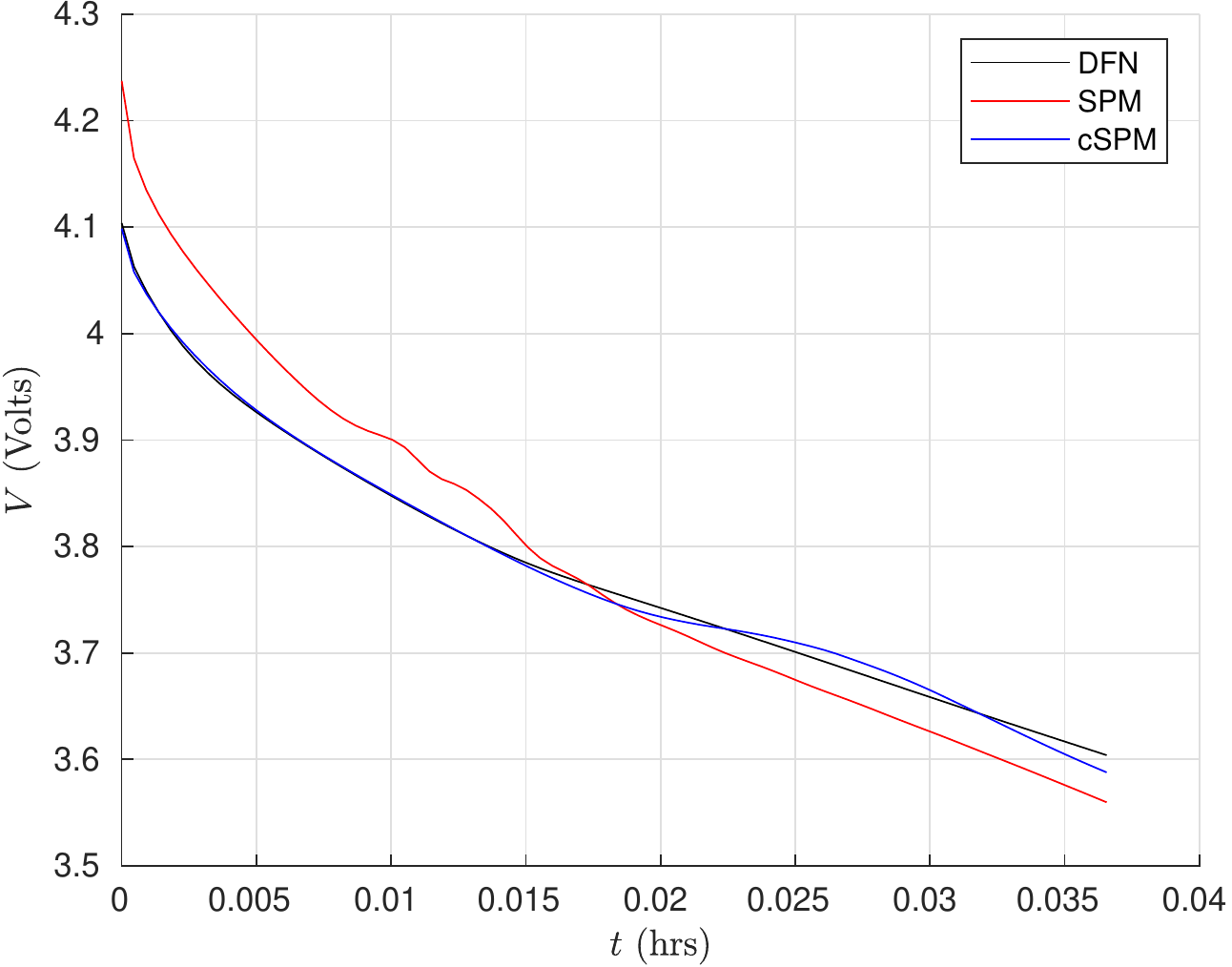}} \qquad
\subfigure[]{\includegraphics[width=0.35\textwidth]{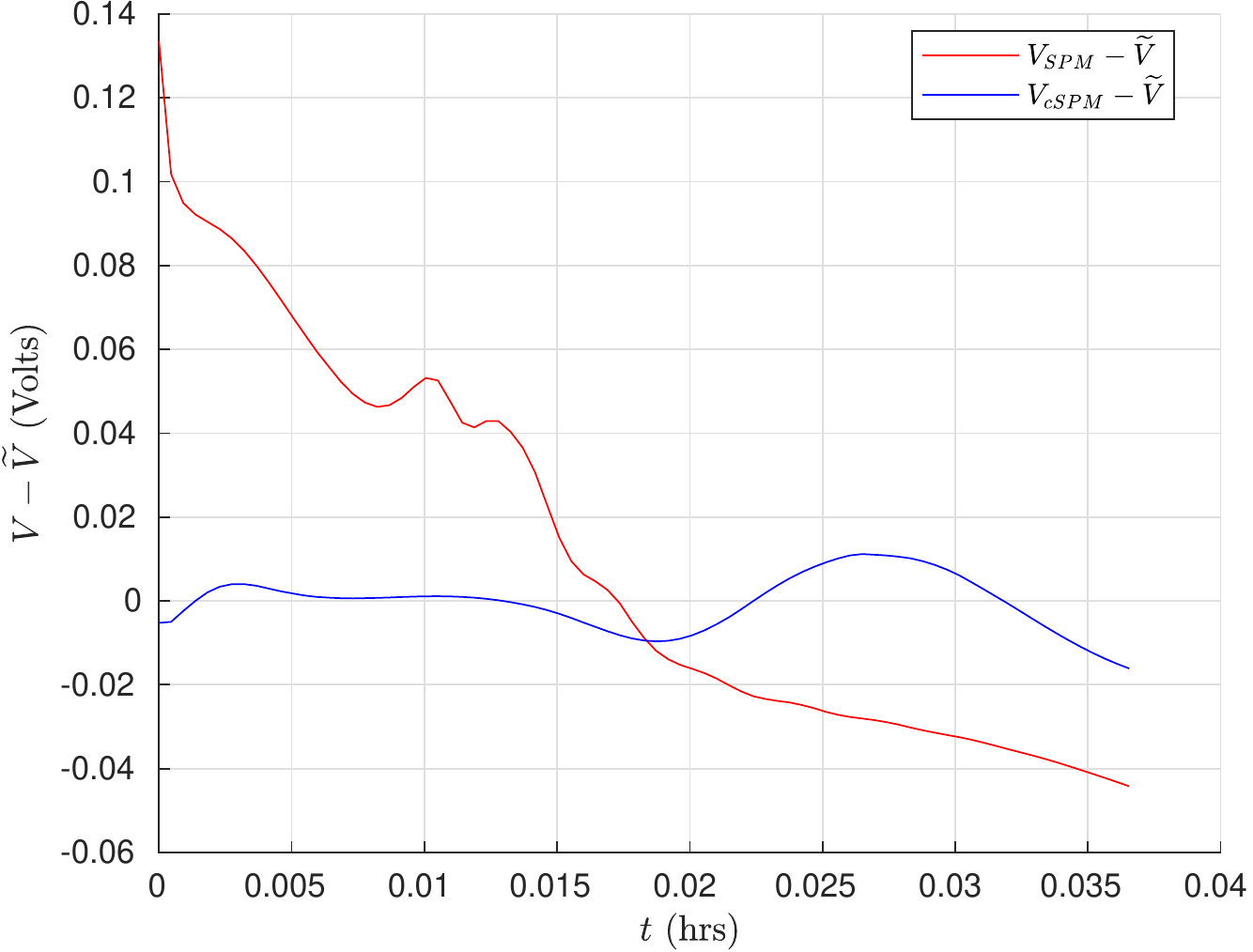}}}
\caption{(Left column) Voltage curves obtained using the DFN model
  with the true parameters $\bP^*$ and predicted by the SP and cSP
  models using optimal values of the reconstructed parameters for
  different C-rates (a)--(b) 2C, (c)--(d) 4C, (e)--(f) 8C, (g)--(h)
  16C. The figures in the right column show the differences $\tV(t) -
  V(t)$ of the voltage curves presented in the left column.}
\label{fig:V}
\end{figure}

To finish the presentation of the results, in Figure \ref{fig:V} we
compare the voltage curves obtained using the DFN model with true
parameter values $\bP^*$ and the voltage curves predicted by the SP
and cSP models with optimal parameter values at different C-rates.
These ``optimal'' sets of parameters are the elements of the Markov
chains shown in Figures \ref{fig:RGB}(a--d) which correspond to the
smallest values of the error functional $\J(\bP)$ (i.e., marked with
the largest symbols in the figures). Figure \ref{fig:V} confirms that
the voltage curves predicted by the cSP model are much closer to the
``true'' voltage curves than those obtained with the SP model,
although the quality of the fits in both cases deteriorates as the
C-rate increases. Moreover, the SP model tends to always overestimate
the true voltage at early stages of the experiment.

\FloatBarrier

\section{Summary and Conclusions}
\label{sec:final}

In this study we have considered the inverse problem of parameterizing
Newman-type models of lithium transport focusing on quantifying the
inherent uncertainty of this process. In order to isolate intrinsic
mechanisms responsible for this uncertainty, we have concentrated on
an idealized problem where ``synthetic'' measurements are manufactured
using the most complete DFN model with parameter values considered
``true'', whereas parametrization is performed based on simplified
versions of this model, namely, the SP and cSP models. By framing the
problem in this way, we are able to eliminate aspects which affect
uncertainty, but are hard to quantity such as, e.g., experimental
errors.

Since the SP model involves one parameter only, the calibration
problem can be solved in this case simply by plotting the error
functional as a function of the unknown parameter $\widehat{D}_s$,
cf.~Figure \ref{fig:SPM}. On the other hand, the cSP model involves
five adjustable material parameters and the corresponding inverse
problem is solved using the Bayesian approach where the reconstructed
parameters are represented in terms of the posterior probability
distributions quantifying the relative uncertainties of different
values.

As a main finding of this study, we reveal an inherent trade-off
between accuracy and uncertainty in the parametrization process.  More
specifically, when the C-rate is small, the two optimally parameterized
simplified models fit the measurement data very well, although good
fits are obtained with a broad range of parameters values thus making
it hard to ascertain precise values of these parameters, cf.~Figures
\ref{fig:SPM}--\ref{fig:PDF}. On the other hand, for large C-rates the
best fits to measurements are obtained with more narrowly determined
parameters, although the accuracy of these fits deteriorates.
Moreover, while they are less uncertain, the parameter values inferred
at higher C-rates tend to have larger errors with respect to the true
values. {Thus, one can conclude that, as C-rate increases,
  uncertainty of inverse modelling is traded for inaccuracy.} These
observations highlight the challenges involved in inferring unknown
material properties via inverse modelling based on simplified models
and voltage curves used as measurements. It is possible that these
difficulties could be mitigated by using measurements of additional
quantities in the parametrization process, such as, e.g., the
concentration of lithium as was done in
\cite{skhgp15a,sethurajan19,ekfkgp20a}, although such measurements are
usually much harder to obtain.

Our overall conclusions are that obtaining parameter values using
simplified versions of the DFN model is a viable and useful strategy
because these simplified models reduce the size of the parameter space
and hence drastically reduce the computational effort required in the
fitting process. However, care must be taken to make sure that the
ranges of validity of the simplified models coincide with the
operating regimes in which the experimental data was harvested. If
this is not the case, we are essentially attempting to fit an invalid
model and accurate results cannot be expected. Finally, higher C-rate
data is more valuable for inverse modelling than low C-rate data
because cell voltages are more strongly depend upon a variety of
parameters (at low C-rates {the cell voltage is almost entirely
  determined by $U_{eq}$}) which allows us to reduce the amount of
uncertainty in their reconstruction.

\section*{Acknowledgments}

JME and BP were supported by a Collaborative Research \& Development
grant \# CRD494074-16 from Natural Sciences \& Engineering Research
Council of Canada. Support from the University of Texas at San Antonio
is also gratefully acknowledged by JME. JF and SS were supported by
the Faraday Institution MultiScale Modelling (MSM) project Grant
number EP/S003053/1.

\appendix

\section{Numerical Approach}
\label{sec:numerics}

{In this appendix we provide some details concerning our approach to
  the numerical solution of the DFN, SP and cSP models, and the
  achieved levels of accuracy. This description focuses on the DFN
  model, but similar, suitably simplified, methods are also used to
  solve the SP and cSP models.}  Our numerical method consist of using
(i) finite element approximation given in \cite{Korotkin21} for
spatial discretization {(in the macroscopic variable $x$) of the}
electrolyte equations \eqref{hc1}--\eqref{hc7}, {and also} for {the
  treatment of }spatial discretization {(in the microscopic variable
  $r$) of the} microscopic equations \eqref{hc8}-\eqref{hc9} and (ii)
and MATLAB's {\tt ode15s} routine for temporal integration.
{Approximation is both {$x$} and $r$ is second-order accurate.}
First, {to discretize the coordinate $x$,} we introduce a set of
points $x_i$, $i=0,\dots,N$ with step size $\Delta_x = x_{i+1}- x_i$
{and denote the corresponding} values of $c(x,t)$, $\Phi(x,t)$ and
$\Phi_s (x,t)$ by {$c_i(t)$, $\Phi_i(t)$ and $\Phi_{s,i}(t)$.}  At
each of these grid points {$x_i$ we need to determine the lithium
  concentration by solving equation \eqref{hc8}.  Discretizing the
  microscopic coordinate $r$ with} $r_j= j\Delta_r$, where $\Delta_r =
1/(M-1)$ for $j=1,\dots,M$, we have $N\times M$ different stations in
$r$.  We denote the values of lithium concentration {at these
  locations by $c_{s,i}^j(t)$, where} the index $i$ represents
particle's position in $x$ whereas $j$ represents radial position
within the particle.

In total, we have $2N$ equations for the concentration and potential
in the electrolyte, $N$ equations for the potential in the solid, and
$M\times N$ equations for the concentration in the solid. Our solution
vector {thus consists of a} total of {$(3+M)\times N$} unknowns. We assemble
the unknown functions of time into one large  vector ${\bf u}(t)$ as
follows
\be\label{sol:vec1}
{\bf u}(t) = [c_1,...,c_N,\phi_1,...,\phi_N,\phi_{s1},...\phi_{sN}, c_{s,1}^1,...,c_{s,N}^1,...,c_{s,1}^M,...c_{s,N}^M]^T\\
=[{\bf c}(t),{\bf \Phi}(t),{\bf \Phi_s}(t),{\bf c_s}(t)]^T
\ee
where the subscript $T$ denotes the transpose. This leads us to the following system of DAEs
\be\label{dae}
{\bf{M}}\frac{d{{\bf u}}}{dt} ={\bf f(u)}, \qquad \text{with} \qquad{\bf u}|_{t=0} ={\bf u}_0.
\ee
Here ${\bf M}_{{(3+M)\times N \times(3+M)\times N }}$ is the mass matrix whose entries
are coefficients of the time derivatives and the function ${\bf f(u)}$
is non-linear and {returns a vector of dimension} {$(3+M)\times N$}. Its
entries {represent} the right-hand side of the discretized
equations. To solve this temporal system of DAEs we use {\tt ode15s}
from MATLAB. 

As we have already mentioned in Section \ref{sec:cSPM}, solution of
the cSP model consists of three steps. In the first step we solve the
1D diffusion equation \eqref{hc8} with boundary condition \eqref{hc9}
and initial condition \eqref{hc11} in a representative electrode
particle, which is the SP model \eqref{spm1}--\eqref{spm3}. In the
second step we solve the electrolyte equations
\eqref{hc1}--\eqref{hc2} with boundary condition \eqref{hc6} and
initial condition \eqref{hc11}. Note that both these system of DAEs
are solved independently. The first step (the SP model) requires a
discretization in $r$ with $M$ grid points and the second step
requires a discretization in $x$ with $N$ grid points. The dimensions
of the DAEs systems that we need to integrate in time for the SP
model, cSP model and DFN model are $M$, $2N+M$ and $(3+M)\times N$,
respectively.  It is clear that the cSP model is computationally more
expensive than the SP model, but it is still very fast compared to the
DFN model. For instance, for $M=50$ and $N=50$ the simulation time is
31 seconds for the DFN model, 0.32 seconds for the cSP model and
$0.12$ seconds for the SP model. All these simulation have been
performed on an Intel(R) Core(TM) i9-9880H CPU @ 2.30GHz.

\begin{figure}[t!]
\centering
\mbox{
\subfigure[]{\includegraphics[width=0.5\linewidth]{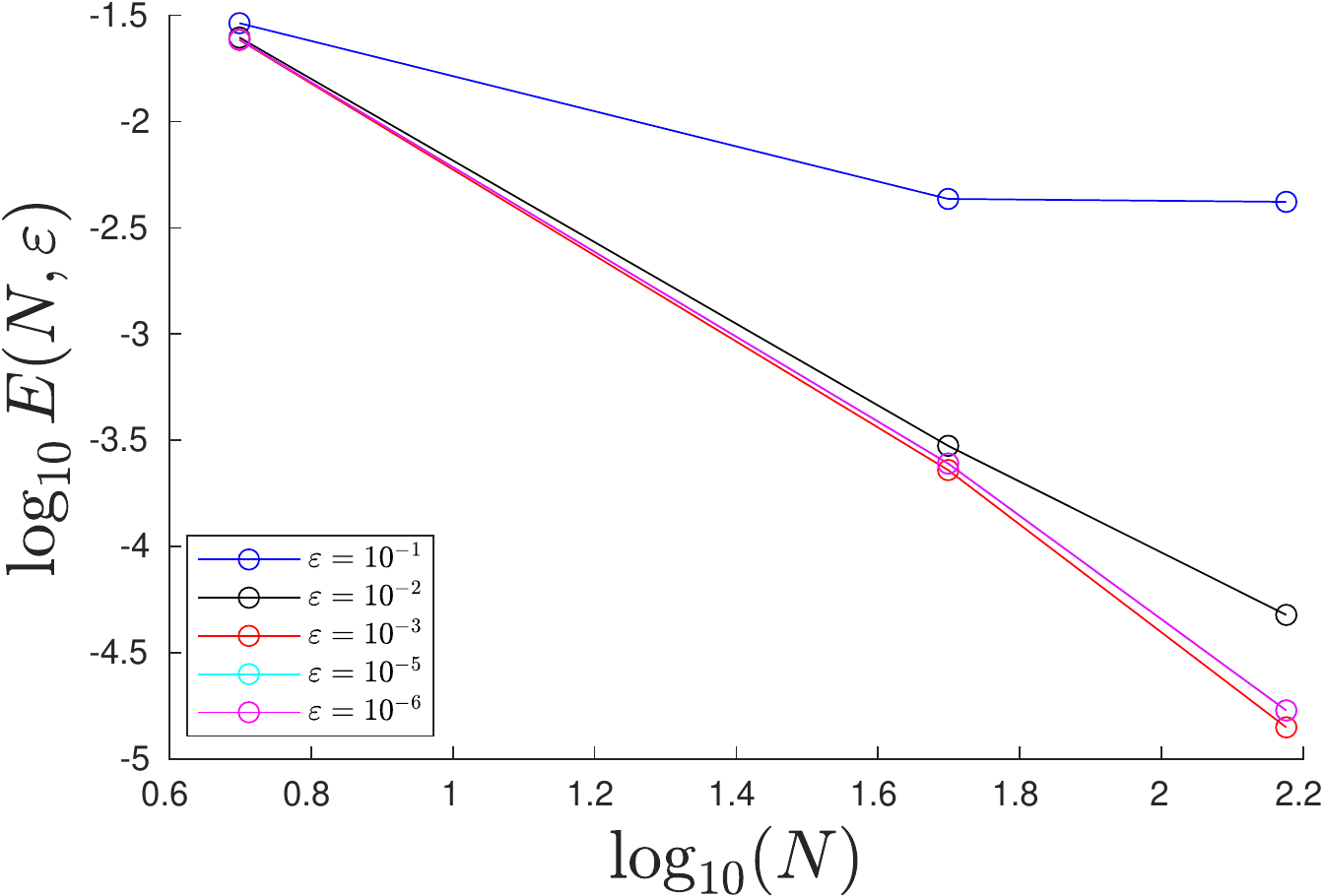}}
\subfigure[]{\includegraphics[width=0.5\linewidth]{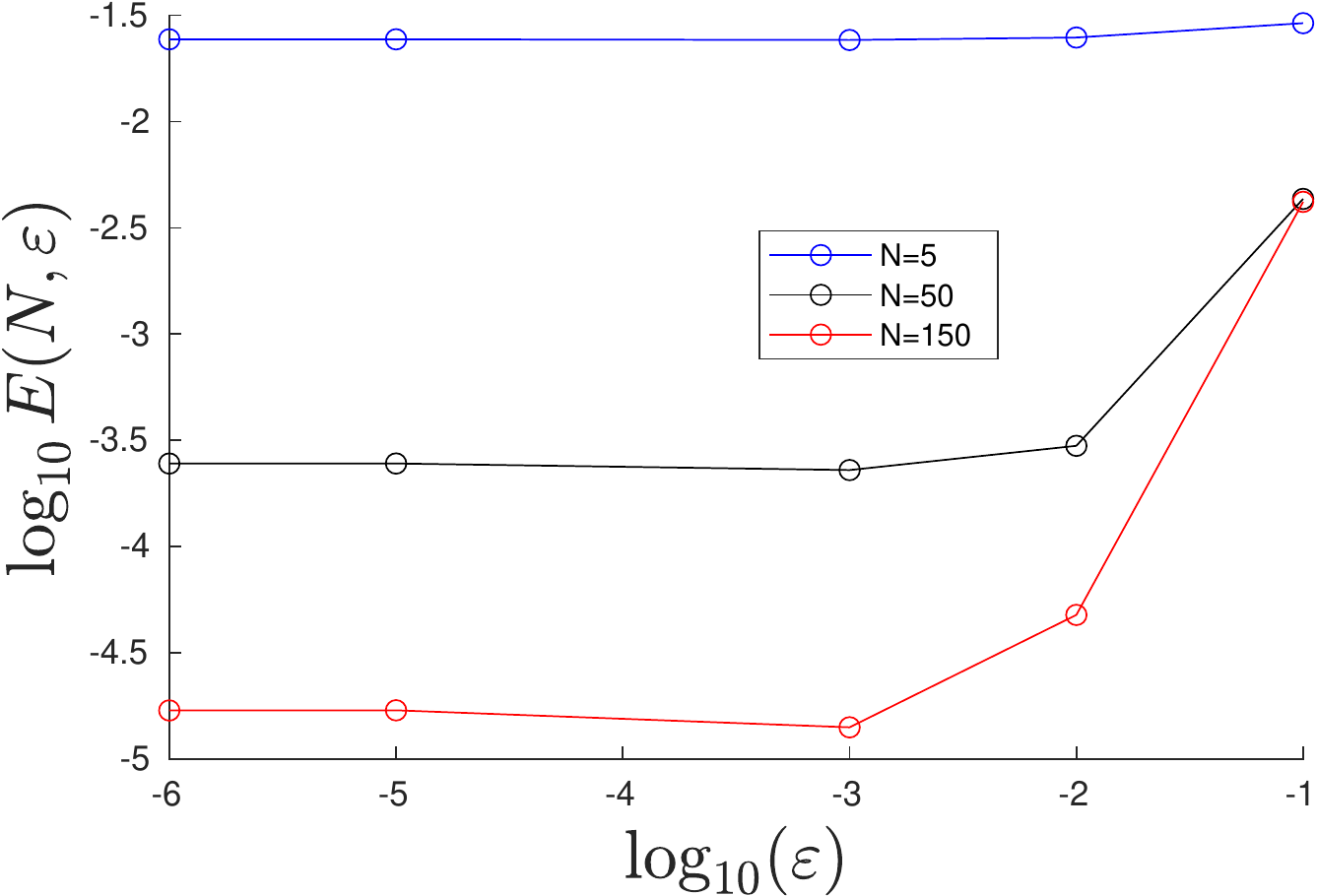}}}
\caption{{Dependence of the mean-square error \eqref{eq:E} on (a) 
    $N$ for different fixed $\varepsilon$ and (b) $\varepsilon$ for
    different fixed $N$.}}
\label{fig:E}
\end{figure}

The accuracy of the numerical solution is determined by the parameters
$M$ and $N$ as well as the tolerance (both relative and absolute) of
time integration $\varepsilon$ (which is used as a parameter by the
function {\tt ode15s}). These errors implicitly determine a lower
bound on the error functional \eqref{eq:J} below which this expression
cannot be reduced, which in turn is needed to properly formulate the
Bayesian inference problem in Section \ref{sec:inverse}.. In order to
estimate how this lower bound depends on the numerical parameters, we
defined the quantity
\begin{equation}
E(N,\varepsilon) = \int_{0}^{t_f} | V_{DFN}(N_0,\varepsilon_0) - V_{DFN}(N,\varepsilon) |^2 \, dt,
\label{eq:E}
\end{equation}
where $V_{DFN}(N,\varepsilon)$ is the voltage obtained by solving
numerically the DFN system \eqref{hc1}--\eqref{hc9} using the
numerical parameters $N$ and $\varepsilon$, whereas $N_0 = 500$ and
$\varepsilon_0 = 10^{-6}$ are the most refined values of these
parameters we consider. Thus, $V_{DFN}(N_0,\varepsilon_0)$ may be
regarded as the ``true'' voltage and expression \eqref{eq:E} as a
mean-square error due to numerical approximation of the DFN system
\eqref{hc1}--\eqref{hc9}. The dependence of $E(N,\varepsilon)$ on $N$
with $\varepsilon$ fixed and on $\varepsilon$ with $N$ fixed in shown
in Figures \ref{fig:E}(a) and \ref{fig:E}(b), respectively. Both these
figures show the expected behavior with error \eqref{eq:E} reduced as
the numerical parameters $N$ and $\varepsilon$ are refined. Based on
this data, we chose to perform our computations with $N = 50$ and
$\varepsilon = 10^{-3}$ such that the corresponding inaccuracy in the
evaluation of the error functional \eqref{eq:J} can be conservatively
estimated as $\mathcal{O}(10^{-3})$. This choice of the numerical
parameters thus balances accuracy with computational cost.

\FloatBarrier

\end{document}